\documentclass{ieeeaccess}
\usepackage{cite}
\usepackage{amsmath,amssymb,amsfonts}
\usepackage{mathrsfs}
\usepackage{algorithm}
\usepackage{algorithmicx}
\usepackage{algpseudocode}
\usepackage{graphicx}
\usepackage{textcomp}
\usepackage[hidelinks]{hyperref}
\usepackage{subfigure}
\usepackage{array}

\def\BibTeX{{\rm B\kern-.05em{\sc i\kern-.025em b}\kern-.08em
    T\kern-.1667em\lower.7ex\hbox{E}\kern-.125emX}}

\begin{document}
\history{Date of publication xxxx 00, 0000, date of current version xxxx 00, 0000.}
\doi{10.1109/ACCESS.2023.0322000}

\title{Intelligent multicast routing method based on multi-agent deep reinforcement learning in SDWN}
\author{\uppercase{HONGWEN HU}\authorrefmark{1}, \uppercase{MIAO YE}\authorrefmark{2,3}, \uppercase{CHENWEI ZHAO}\authorrefmark{3}, \uppercase{QIUXIANG JIANG}\authorrefmark{2,3}, \uppercase{YONG WANG}\authorrefmark{1}, \uppercase{HONGBING QIU}\authorrefmark{3} and \uppercase{XIAOFANG DENG}\authorrefmark{3}}

\address[1]{School of Computer Science and Information Security, Guilin University of Electronic Technology, Guilin 541004, China}
\address[2]{Guangxi Key Laboratory of Wireless Wideband Communication and Signal Processing, Guilin University of Electronic Technology, Guilin 541004, China}
\address[3]{School of Information and Communication, Guilin University of Electronic Technology, Guilin 541004, China}
\tfootnote{This work was supported in part by the National Natural Science Foundation of China (Nos.62161006, 62172095, 61861013), the subsidization of the Innovation Project of Guangxi Graduate Education (Nos. YCSW2022270, YCSW2023310, YCBZ2023134), and Guangxi Key Laboratory of Wireless Wideband Communication and Signal Processing (No. GXKL06220110).}

\markboth
{Author \headeretal: Preparation of Papers for IEEE TRANSACTIONS and JOURNALS}
{Author \headeretal: Preparation of Papers for IEEE TRANSACTIONS and JOURNALS}

\corresp{Corresponding author: Xiaofang Deng (e-mail: xfdeng@guet.edu.cn)}

\begin{abstract}
Multicast communication technology is widely applied in wireless environments with a high device density. Traditional wireless network architectures have difficulty flexibly obtaining and maintaining global network state information and cannot quickly respond to network state changes, thus affecting the throughput, delay, and other QoS requirements of existing multicasting solutions. Therefore, this paper proposes a new multicast routing method based on multiagent deep reinforcement learning (MADRL-MR) in a software-defined wireless networking (SDWN) environment. First, SDWN technology is adopted to flexibly configure the network and obtain network state information in the form of traffic matrices representing global network links information, such as link bandwidth, delay, and packet loss rate. Second, the multicast routing problem is divided into multiple subproblems, which are solved through multiagent cooperation. To enable each agent to accurately understand the current network state and the status of multicast tree construction, the state space of each agent is designed based on the traffic and multicast tree status matrices, and the set of AP nodes in the network is used as the action space. A novel single-hop action strategy is designed, along with a reward function based on the four states that may occur during tree construction: progress, invalid, loop, and termination. Finally, a decentralized training approach is combined with transfer learning to enable each agent to quickly adapt to dynamic network changes and accelerate convergence. Simulation experiments show that MADRL-MR outperforms existing algorithms in terms of throughput, delay, packet loss rate, etc., and can establish more intelligent multicast routes. 
\end{abstract}


\titlepgskip=-21pt

\maketitle

\section{Introduction}
\label{sec:introduction}
\PARstart{W}{ith} the rapid development of wireless network technology, the applications of multicast communication in wireless networks are becoming increasingly widespread. This communication technology can be applied for purposes such as video live streaming, multimedia conferences, real-time data transmission, and online games. In these applications, as the number of users and the level of user demand continue to increase, attempting to use unicast communication to send the necessary data would place enormous pressure on the information sources and the network bandwidth, leading to network congestion and inability to meet user needs. In the broadcasting scenario, the transmitted information will also be received by users who do not need it, not only compromising the security of the information but also wasting considerable bandwidth. For such point-to-point applications, multicast technology can better solve the above problems. In multicast services, only one multicast message needs to be sent by the source host, and the data are then replicated and distributed to multiple target nodes upon encountering forked nodes during transmission \cite{b1}. Therefore, multicasting can effectively save bandwidth, reduce the network load, and improve the security of information transmission \cite{b2}.

Multicast routing requires the construction of an optimal multicast tree from the source node to all destination nodes \cite{b3}. Timely acquisition of global dynamic network state information is one of the basic prerequisites for constructing such an optimal multicast tree. Traditional wireless networks typically utilize a distributed management approach \cite{b4}, in which network resources and functionalities are dispersed across various wireless network devices (such as access points, routers, and switches) and each device independently executes control decisions. While this approach offers flexibility, it suffers from low management efficiency and presents difficulties in achieving timely optimization and coordination of the entire network. Additionally, as the network expands in scale, the traffic data forwarded by network devices become increasingly voluminous, making it challenging for traditional network devices, for which forwarding is tightly coupled with control, to obtain real-time information on the global network status. To address the aforementioned issues, the recently emerging technology of software-defined wireless networking (SDWN) \cite{b5} provides an excellent solution.

SDWN combines software-defined networking (SDN) \cite{b6} with wireless networks. SDWN solves the problems of low network management and control efficiency and the difficulty of achieving global optimization and coordination in traditional wireless network structures by exploiting the centralized management advantages of SDN, such as centralized control logic and the decoupling of forwarding from control. By taking full advantage of these centralized management capabilities, SDWN facilitates the global optimization and coordination of network resources. SDWN enables the controller of a wireless network to obtain the global static topological structure of the network, the global network state, and the utilization rates of resources by controlling the logical concentration \cite{b7}. In combination with the programmability of SDWN networks, these capabilities allow the network controller to achieve unified management, integration, and virtualization of network resources and to use a northbound interface to provide on-demand allocation of network resources and services for upper-layer applications.

The classic algorithms for constructing multicast trees in traditional multicast routing include the shortest path and minimum spanning tree algorithm of Kou, Markowsky and Berman (the KMB algorithm) \cite{b8}, the minimum cost path heuristic (MPH) algorithm \cite{b9}, and the average distance heuristic (ADH) algorithm \cite{b10}. These classic multicast tree construction algorithms have been successfully applied in many fields over the past decade. However, with the continuous expansion of the network scale and the exponential growth in network traffic, these traditional methods cannot adapt to the dynamic and changing wireless network environment, making it difficult to meet the current requirements in terms of network service quality. Moreover, as the scale of SDWN networks continues to expand, this deficiency becomes even more apparent. Therefore, designing multicast trees that can adapt to dynamic network characteristics by exploiting the advantages of the SDN architecture in order to meet the high-performance requirements of multicast services is an important research topic.

In recent years, artificial intelligence technology has been increasingly studied and applied in the networking field due to its strong adaptability and flexibility. Deep reinforcement learning has significant advantages in high-dimensional and complex decision-making. By combining it with the SDN architecture, researchers can fully leverage its flexibility and ability to adapt to network dynamics, thereby improving network efficiency and performance. Currently, most research on the application of deep reinforcement learning to SDN unicast and multicast communication is limited to discussions of single-agent reinforcement learning methods \cite{b11,b12,b13,b14} However, compared to multi-agent reinforcement learning, the convergence speed of these methods is slow. Consequently, in the case of frequent and dynamic changes in the network state, the single-agent approach has difficulty responding quickly to the forwarding needs of data flows.

In consideration of the above issues, this paper proposes an intelligent multicast routing method based on multiagent deep reinforcement learning, named MADRL-MR, for use in SDWN. In MADRL-MR, an SDWN framework is designed to overcome the limitations of traditional wireless networking, in which the overall network cannot be directly controlled and maintained, and to enable more convenient configuration of the network devices while improving the network performance. This framework is used to manage a wireless network and obtain its global topology and link state information. It also makes use of the adaptability and flexibility of deep reinforcement learning to adapt to dynamic changes in the network. To address the slow convergence speed of the construction of multicast trees using a single intelligent agent as well as the difficulty of quickly responding to data forwarding demands, a multi-agent deep reinforcement learning algorithm is designed for multicast tree construction in MADRL-MR. In this algorithm, each intelligent agent can independently learn and adapt to changes in the network state and collaborate to achieve better routing strategies. To accelerate the training speed of the multiple intelligent agents, we design corresponding transfer learning mechanisms \cite{b15}, in which an initial set of weights is pre-trained and loaded before each intelligent agent begins training to accelerate its convergence speed.

The main contributions of this article are as follows:

\begin{enumerate}
\item In contrast to the traditional approach for managing and maintaining the global network state in a wireless network, we design a network architecture based on SDWN. By virtue of the centralized control logic and programmability features of SDWN, we can monitor the global static topology and network status information of a wireless network and obtain real-time link status information, such as bandwidth, delay, and packet loss rate, to achieve more efficient global optimization and coordination of the network resources.

\item In contrast to the existing method of building multicast trees with a single intelligent agent, we design and implement an intelligent multicast routing method based on multiagent deep reinforcement learning. First, we divide the problem of multicast tree construction into multiple subproblems, which are solved through collaboration among multiple intelligent agents. Second, in the design of the state space for each intelligent agent, we comprehensively consider parameters such as bandwidth, delay, the packet loss rate of wireless links, the used bandwidth, the packet error rate, the packet drop rate, the distance between access points, and the multicast tree construction status. In addition, instead of the existing method of using the k-paths approach to design the action space for an intelligent agent, we design a novel action space using the next-hop node in the network as the action. Finally, we design corresponding reward functions for the four possible scenarios encountered in multicast tree construction, which can guide the intelligent agents to select efficient multicast routes.

\item To improve the convergence efficiency and collaboration stability of the multiple intelligent agents, we design a fully decentralized training (independent learning, IL) method for multiagent systems. In addition, to enhance the convergence speed of the multiagent system, we adopt transfer learning techniques. Specifically, we transfer knowledge acquired from experts or previous tasks to the current task at the beginning of the training process, thereby reducing the initial ineffective exploration of the intelligent agents and accelerating their convergence.
\end{enumerate}

The rest of this article is organized as follows. Section \ref{sec:related work} introduces the relevant work. Section \ref{sec:DESIGN OF SYSTEM ARCHITECTURE} analyzes the problem and introduces the SDWN intelligent multicast routing structure. Section \ref{sec:MADRL-MR algorithm} provides a detailed introduction to the MADRL-MR algorithm. Section \ref{sec:Experiment} introduces the experimental setup and performance evaluation results. Section \ref{sec:Conclusion} introduces the conclusion and future work.

\section{RELATED WORK}
\label{sec:related work}
In this section, we mainly discuss the related work on multicast routing in SDWN and analyze the advantages and disadvantages of traditional algorithms and intelligent algorithms applied in multicast routing.

\textsl{\textbf{Traditional algorithms:}} Kou et al. \cite{b8} proposed a Steiner tree construction method based on a shortest path and minimum spanning tree algorithm (the KMB algorithm). Takahashi et al. \cite{b9} proposed the minimum cost path heuristic (MPH) algorithm. Smith et al. \cite{b10} designed an algorithm based on an average distance heuristic (ADH). The above three classic algorithms were initially proposed to solve the problem of constructing multicast trees, and many subsequent improvements have been developed based on these algorithms. Yu et al. \cite{b16} proposed an improved algorithm based on key nodes (KBMPH) by prioritizing the paths for certain key nodes. Zhou et al. \cite{b17} designed a delay-constrained MPH algorithm (DCMPH). Zhao et al. \cite{b18} studied how to reduce the cost of constructing a Steiner tree and proposed a weighted node-based MPH algorithm (NWMPH). Farzinvash et al. \cite{b19} decomposed the problem of multicast tree construction in a wireless mesh network into two phases, with the first phase considering delay and the second phase considering bandwidth. By combining the two phases, these authors proposed an algorithm that comprehensively considers both delay and bandwidth for the construction of multicast trees. Przewoźniczek et al. \cite{b20} transformed k-shortest Steiner tree problems into binary dynamic problems and solved them using the integer linear programming (ILP) method. Walkowiak et al. \cite{b21} used a unicast path construction method to construct a multicast tree, but its computational cost was too high. Martins et al. \cite{b22} transformed the multicast tree construction problem into an ILP problem and designed a heuristic algorithm with delay constraints. Zhang et al. \cite{b23} proposed a delay-optimized multicast routing scheme for use in the SDN context, which utilizes SDN to obtain network state information. Hu et al. \cite{b24} also proposed a multicast routing method based on SDN. However, the traditional algorithms mentioned above can use only a single network resource to construct a multicast tree; thus, they have poor perception of dynamic network changes and significant limitations in constructing efficient multicast routes.

\textsl{\textbf{Intelligent Algorithms:}} Annapurna et al. \cite{b25} proposed a Steiner tree construction method based on ant colony optimization (ACO), which optimizes the Steiner tree using bandwidth, delay, and path cost. Zhang et al. \cite{b26} proposed a multicast routing method based on a hybrid ant colony algorithm. This method combines the solution generation process of the ACO algorithm with the cloud model (CM) to obtain a minimum-cost multicast tree that satisfies bandwidth, delay, and delay jitter constraints. Zhang et al. \cite{b27} proposed a Steiner tree construction method based on particle swarm optimization (PSO), which uses the Steiner tree length as the constraint condition. Nath et al. \cite{b28} used gradient descent based on general PSO to accelerate the convergence speed of PSO and designed a gradient-based PSO algorithm for building a Steiner tree. Zhang et al. \cite{b29} proposed a multicast routing method based on a genetic algorithm (GA), in which a new crossover mechanism called leaf crossing (LC) is introduced into the GA to solve multicast quality of service (QoS) models. The above algorithms are all designed for application in traditional network structures and can use only limited network resources to construct multicast trees. Moreover, these algorithms have high computational complexity and consume a significant amount of time; thus, they have difficulty reaching convergence.

\textsl{\textbf{Reinforcement learning algorithms:}} Heo et al. \cite{b30} proposed a multicast tree construction technique based on reinforcement learning for use in an SDN environment. This technique abstracts the process of constructing a multicast tree as a Markov decision process (MDP), uses SDN technology to obtain global network information and applies reinforcement learning for multicast tree construction. However, this method considers only the number of hops and does not consider other network link state information. Araqi et al. \cite{b31} proposed a Q-learning-based multicast routing method for wireless mesh networks, which considers only channel selection and rate and does not optimize the construction of multicast trees. Tran et al. \cite{b32} proposed a deep Q-network (DQN)-based multicast routing method. In this method, broadcasting is first used to find the destination node, and the destination node then uses unicast communication to send data packets to the source node to complete the construction of the multicast tree. This method only considers delay and does not consider parameters such as bandwidth and packet loss rate. Chae et al. \cite{b33} proposed a multicast tree construction algorithm based on meta-reinforcement learning for use in the SDN context. This algorithm sets the link cost to a fixed value of 1 and does not consider changes in the link state. Zhao et al. \cite{b34} designed a deep reinforcement learning method for intelligent multicast routing in SDN based on a DQN, which considers only the bandwidth, delay, and packet loss rate of each link; this method has the problem of slow convergence of the intelligent agent.

\textsl{\textbf{Multi-agent reinforcement learning algorithms:}} At present, there is still little research in the literature on the application of multiagent deep reinforcement learning methods to multicast problems in wireless networks. Instead, we can refer only to other relevant literature on multiagent deep reinforcement learning algorithms. Yang et al. \cite{b35} proposed a software-defined urban traffic control algorithm based on multiagent deep reinforcement learning for use in a software-defined Internet of Things (SD-IoT) cooperative traffic light environment. Suzuki et al. \cite{b36} proposed a dynamic virtual network (VN) allocation method based on collaborative multiagent deep reinforcement learning (Coop-MADRL) to maximize the utilization of limited network resources in dynamic VNs. Wu et al. \cite{b37} designed a flow control and multichannel reallocation (TCCA-MADDPG) algorithm based on a multiagent deep deterministic policy gradient (MADDPG) algorithm to optimize the multichannel reallocation framework of the core backbone network based on flow control in the SDN-IoT. Bhavanasi et al. \cite{b38} proposed a graph convolutional network routing and deep reinforcement learning algorithm for agents, which regards the routing problem as a reinforcement learning problem with two new modifications. Duke et al. \cite{b39} designed a multiagent reinforcement learning framework for transient load detection and prevention in the SDN-IoT. This framework establishes one agent for multipath routing optimization and another agent for malicious DDoS traffic detection and prevention in the network, with the two agents collaborating in the same environment. Typically, similar multiagent algorithms have certain instability issues, which can result in unstable training and difficulty in convergence during the training phase. Therefore, some researchers have applied transfer learning in combination with multiagent deep reinforcement learning.

\textsl{\textbf{Transfer reinforcement learning:}} Torrey et al. \cite{b40} incorporated transfer learning into multiagent reinforcement learning by proposing a teacher–student framework for reinforcement learning. First, an agent is trained as a teacher agent. Then, when training a second student agent for the same task, the fixed policy of the teacher agent can provide suggestions to speed up the learning process. Parisotto et al. \cite{b41} defined a method of multitasking and transfer learning in deep multitasking and reinforcement learning, which guides agents to take actions in different tasks through expert experience and thus accelerates the learning speed of the agents. Silva et al. \cite{b42} proposed a multiagent recommendation framework in which multiple agents can advise each other while learning in a shared environment.

Considering the limitations of classical heuristic algorithms for multicast routing in wireless networks, the computational complexity of intelligent algorithms, and the slow convergence speed of reinforcement learning, we draw inspiration from a previous study of multicast routing in wired networks \cite{b34}. To adapt to dynamic changes in the wireless network traffic while meeting QoS requirements, this paper proposes the adoption of SDWN technology to perceive global network information and designs a multi-agent based deep reinforcement learning algorithm for the construction of multicast trees. This algorithm can overcome the shortcomings of traditional wireless networks in regard to the inability to directly control and maintain the global network and solves the problem of slow convergence of single-agent multicast tree construction methods.

\section{DESIGN OF SDWN INTELLIGENT MULTICAST ROUTING SYSTEM ARCHITECTURE}
\label{sec:DESIGN OF SYSTEM ARCHITECTURE}
\subsection{Multicast Problem Description}
\label{sec:3.1}
\begin{figure}[htbp]
	\centering
	\includegraphics[width=0.4\textwidth]{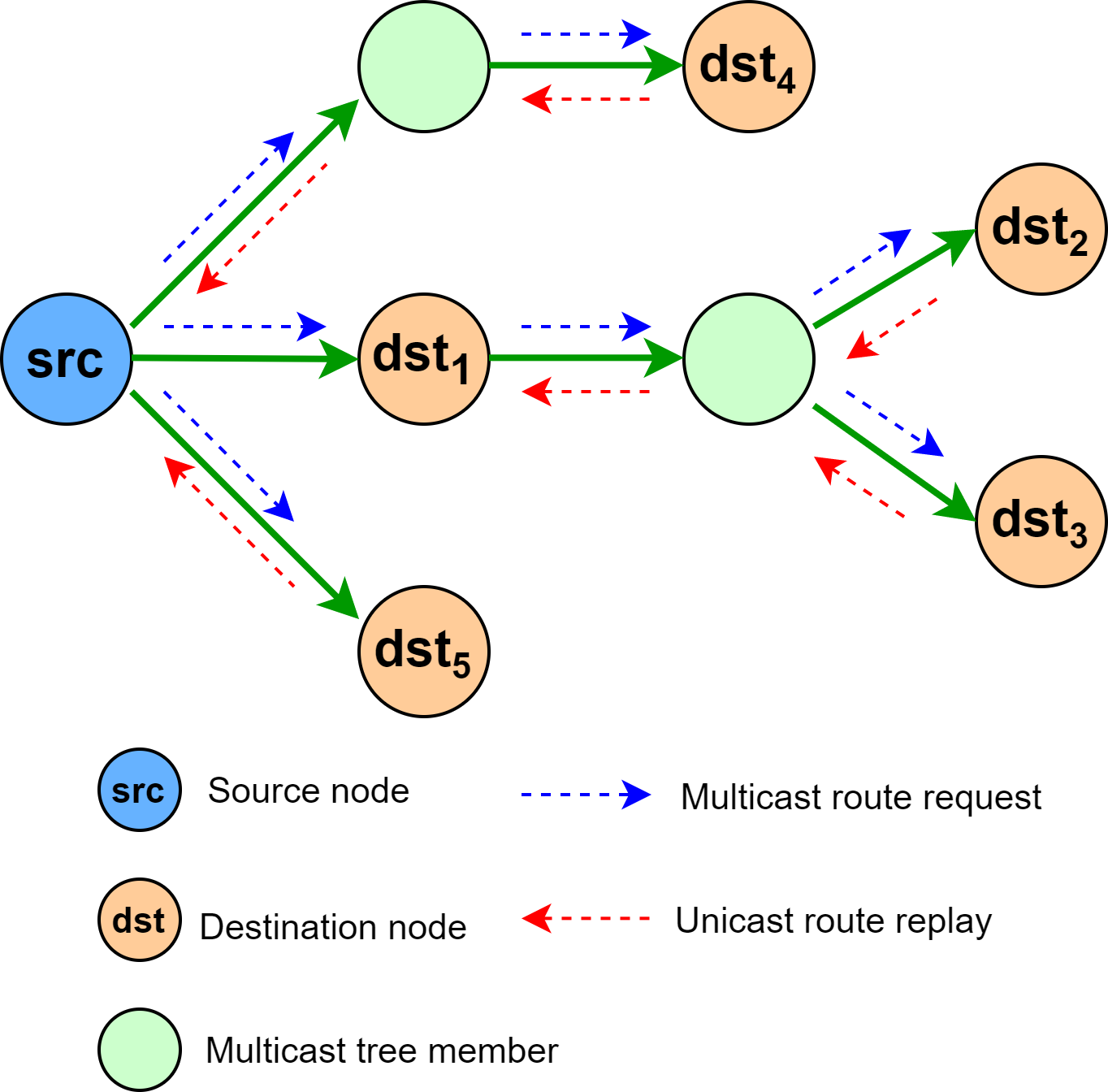}
	\caption{\raggedright Illustrate of data flow directions in a multicast tree}
	\label{fig1}
\end{figure}

Multicast communication, also known as multi-unicast communication, multipoint delivery, or group communication, allows information to be simultaneously transmitted to a group of specified destination addresses. Multicast datagrams are transmitted only once on a link in a network's transport layer and are only duplicated when encountering a branching link. The data flow diagram of multicast network communication is shown in Fig. \ref{fig1}. The data flows in multicast network communication follow a tree-shaped structure called a multicast tree (or Steiner tree), where the source node src is the root of the tree and the destination nodes dst for multicasting are the leaf nodes of the tree. The optimization objective for multicast routing is to find a multicast tree that can achieve the optimal performance.

The optimal multicast tree corresponds to the solution of the mathematically defined Steiner tree problem, which is a classic nondeterministic polynomial time (NP)-complete problem \cite{b9}. Consider a weighted undirected connected graph represented by $G(V,E,w)$, where $V$ is the set of nodes, $E$ is the set of edges, and $w$ specifies the weights of the edges. The edge $e_{ij} \in E $ between node $i$ and node $j$ in the graph has a weight $w(e_{ij})$. Given a subset of nodes $S \subseteq V$, where $S$ contains the source node $src$ and a set of destination nodes $DST=\{dst_1,dst_2,\cdots,dst_n \}$ for multicasting, the nodes in $S$ are the source and destinations of the multicast communication. The graph $G'$ is a subgraph of the graph $ G $ that includes the vertex set $S$. Additionally, $G'$ contains some nodes that are not in the vertex set $S$, which are referred to as Steiner nodes. The objective of the optimal Steiner tree problem is to find a minimum-weight spanning tree $T$ in the graph $G'$ that contains all of the nodes in $S$, as shown in Equation \eqref{eq1}.

\begin{equation}
\mathop {\min }\limits_{T \subseteq G,S \subseteq {V_T}} {\mkern 1mu} \sum\limits_{{e_{ij}}} {w({e_{ij}})}
\label{eq1}
 \end{equation}

Strictly speaking, obtaining an exact optimal solution for this NP-complete multicast tree problem is extremely difficult. Existing works have discussed how to obtain an approximately optimal solution. Accordingly, an approximate treatment can be applied by decomposing the problem into a set of distinct routes from the source node to the multiple destination nodes, as shown in Equation \eqref{eq2}.

\begin{equation}\begin{array}{c}
T = \left\{ {pat{h_1}\left( {src,ds{t_1}} \right), \cdots ,pat{h_k}\left( {src,ds{t_1}} \right),} \right.\\
\left. { \cdots ,pat{h_n}\left( {src,ds{t_n}} \right)} \right\}
\end{array}
\label{eq2}
\end{equation}
where $ path_k $ is the path from the source node $ src $ to $ dst_k $ in the multicast tree $ T $, for $ k=1,2,\cdots,n $, $ dst_k $ belongs to the destination node set $ DST $ of the multicast tree; and $ n $ is the number of destination nodes.

If each $ path_k \in T $ has the minimum cost, then the multicast tree $ T $ is an end-to-end minimum cost tree. Such a multicast tree can be built by constructing each $ path_k $ $ (src,dst_k ) $ as a unicast path and then combining these paths and removing redundant links. Taking advantage of the ability to use SDN technology to monitor the global network resources, this paper calculates the minimum cost $ f(path_k) $ for each path using the following parameters:

$ bw_k $ is the residual bandwidth of $ path_k $, which is the minimum residual bandwidth from the source node $ src $ to the destination node $ dst_k $. Its definition is given in Equation \eqref{eq3}.

\begin{equation}
b{w_k} = \mathop {\min }\limits_{{e_{ij}} \in pat{h_k}} \left( {b{w_{ij}}} \right)
\label{eq3}
\end{equation}
where $ bw_{ij} $ is the remaining bandwidth of the link $ e_{ij} $ between node $ i $ and node $ j $.

$ delay_k $ is the total delay on $ path_k $, which is expressed as the sum of the delays on all links in $ path_k $. Its definition is given in Equation \eqref{eq4}.

\begin{equation}
dela{y_k} = \sum\limits_{{e_{ij}} \in pat{h_k}} {dela{y_{ij}}}
\label{eq4}
\end{equation}
where $ delay_{ij} $ is the delay on the link $ e_{ij} $ between node $ i $ and node $ j $.

$ loss_k $ is the packet loss rate on $ path_k $, which is calculated as shown in Equation \eqref{eq5} since the packet loss rate on some links is 0.

\begin{equation}
los{s_k} = 1 - \prod\limits_{{e_{ij}} \in pat{h_k}} {\left( {1 - los{s_{ij}}} \right)}
\label{eq5}
\end{equation}
where $ loss_{ij} $ is the packet loss rate on the link $ e_{ij} $ between node $ i $ and node $ j $.

$ used\_bw_k $ is the bandwidth used on $ path_k $, which is expressed as the maximum bandwidth used from the source node $ src $ to the destination node $ dst_k $. It is defined as shown in Equation \eqref{eq6}.

\begin{equation}
used\_b{w_k} = \mathop {\max }\limits_{{e_{ij}} \in pat{h_k}} \left( {used\_b{w_{ij}}} \right)
\label{eq6}
\end{equation}
where $ used\_bw_{ij} $ is the bandwidth used on the link $ e_{ij} $ between node $ i $ and node $ j $.

$ errors_k $ is the error packet rate on $ path_k $, which is calculated via Equation \eqref{eq7}.
\begin{equation}
error{s_k} = 1 - \prod\limits_{{e_{ij}} \in pat{h_k}} {\left( {1 - error{s_{ij}}} \right)}
\label{eq7}
\end{equation}
where $ errors_{ij} $ is the packet error rate on the link $ e_{ij} $ between node $ i $ and node $ j $.

$ drops_k $ is the drop rate on $ path_k $, which is calculated via Equation \eqref{eq8}.
\begin{equation}
drop{s_k} = 1 - \prod\limits_{{e_{ij}} \in pat{h_k}} {\left( {1 - drop{s_{ij}}} \right)}
\label{eq8}
\end{equation}
where $ drops_{ij} $ is the packet drop rate on the link $ e_{ij} $ between node $ i $ and node $ j $.

$ distance_k $  is the average distance of each link in $ path_k $. In a wireless network, the distance between access points (APs) will affect data forwarding. The average distance can be used to measure the average energy consumed by each AP node to send data and is defined in Equation \eqref{eq9} below.

\begin{equation}
distanc{e_k} = average\left( {\sum\limits_{{e_{ij}} \in pat{h_k}} {distanc{e_{ij}}} } \right)
\label{eq9}
\end{equation}
where $ distance_{ij} $ is the distance of the link $ e_{ij} $ between node $ i $ and node $ j $.

The objective function  $ f(path_k) $ is formulated to maximize the residual bandwidth $ bw_k $ and minimize the delay $ delay_k $, the packet loss rate $ loss_k $, the used bandwidth $ used\_bw_k $, the packet error rate $ errors_k $, the packet drop rate $ drops_k $ and the average distance $ distance_k $ between wireless APs, as shown in the Equation \eqref{eq10}.

\begin{equation}
\begin{array}{c}
f\left( {pat{h_k}} \right) = {\beta _1}b{w_k} + {\beta _2}los{s_k} + {\beta _3}dela{y_k}\\
\\
+ {\beta _4}used\_b{w_k} + {\beta _5}error{s_k}\\
\\
+ {\beta _6}dro{p_k} + {\beta _7}distanc{e_k}
\end{array}
\label{eq10}
\end{equation}
where $ \beta_l $ represents the weight of parameter $l$, for $ l=1,2,\cdots,7 $. The specific design of $ \beta_l $ is described in section \ref{sec:4.1}, which discusses the reward function design.

The optimization objective value on each path is represented by $ f(path_k) $, and the process of constructing the multicast tree consists of finding such a path for each destination node. These tasks are independent of each other, so the problem of multicast tree construction can be mathematically expressed as the multi-objective optimization problem shown in Equation \eqref{eq11}.

\begin{equation}
\begin{array}{c}
\max F\left( T \right) = \left[ {f\left( {pat{h_1}} \right), \cdots ,f\left( {pat{h_k}} \right),} \right.\\
\\
\left. { \cdots f\left( {pat{h_n}} \right)} \right]
\end{array}
\label{eq11}
\end{equation}
where $ T $ is the multicast tree that implements the communication path of the multicast network, $ path_k $ is the optimal path for each destination node, and $ path_k \in T $, $ n $ is the number of destination nodes.

\subsection{SDWN intelligent multicast routing architecture}
\label{sec:3.2}
The SDWN-based intelligent multicast routing strategy combines SDN with wireless networking, using multiagent reinforcement learning to achieve multicast routing. By perceiving the network state information of the wireless network, we obtain information such as the bandwidth, delay, packet loss rate, used bandwidth, packet error rate, packet drop rate, and distance between wireless access nodes in the wireless network. We use multiagent collaboration to construct multicast paths from the source node to all destination nodes and use the southbound interface of the centralized controller to issue flow tables to the switches on the paths to achieve multicast routing. With its ability to monitor the global network link state information, SDWN enables the agents to intelligently adjust these multicast routes based on dynamic changes in the network state information.

\begin{figure}[htbp]
	\centering
	\includegraphics[width=0.45\textwidth]{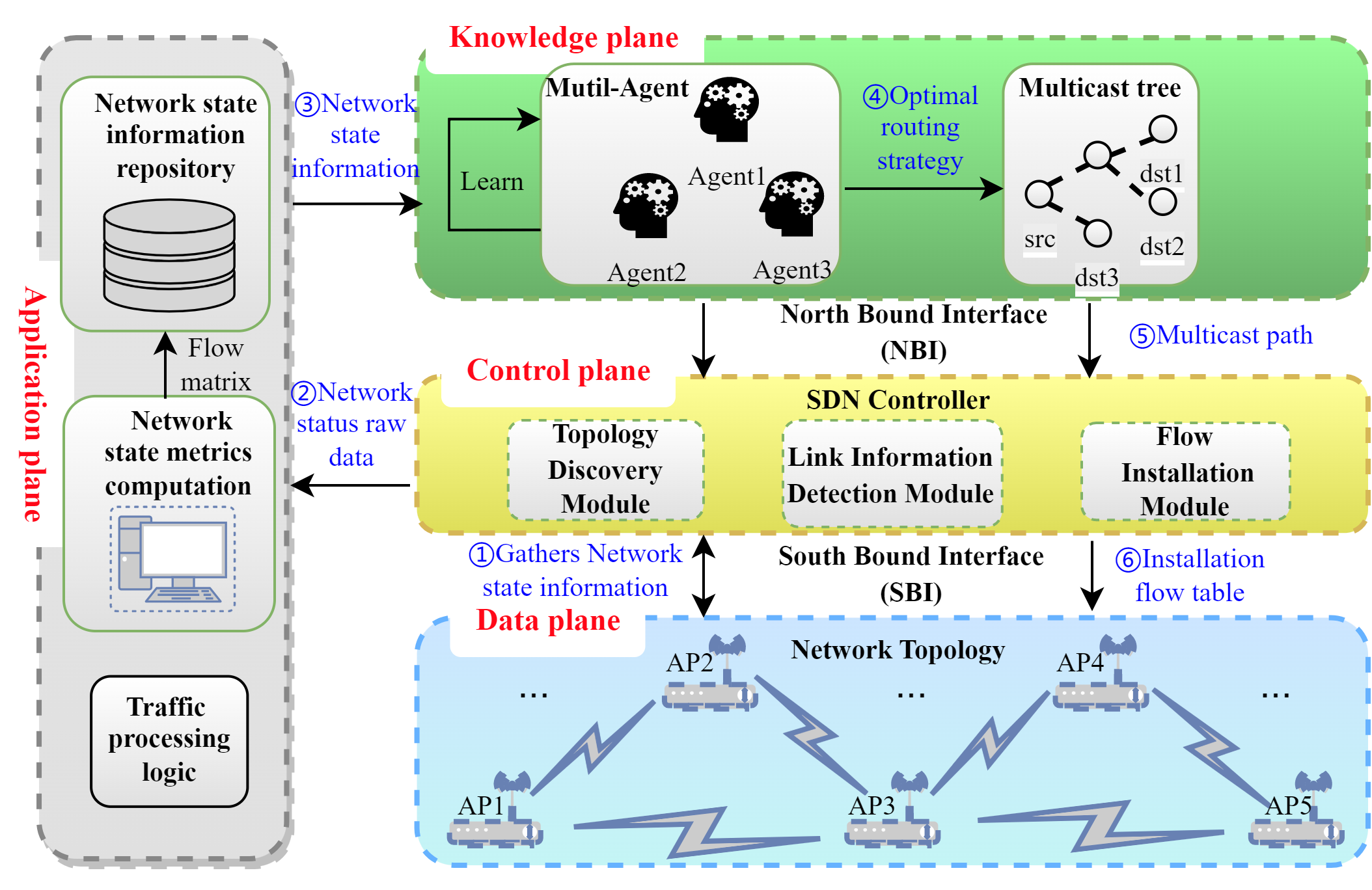}
	\caption{\raggedright SDWN-based intelligent multicast routing architecture}
	\label{fig2}
\end{figure}

The overall structure of the SDWN-based intelligent multicast routing strategy is shown in Fig. \ref{fig2}, and it is explained in further detail below.

\begin{enumerate}
	\renewcommand{\labelenumi}{\textcircled{\scriptsize\theenumi}}
	\item The control plane periodically retrieves network status information from the data plane.
	\item The application plane collects raw data on the network status and processes these data into corresponding traffic matrices.
	\item The knowledge plane utilizes the processed traffic matrices from the application plane.
	\item Each intelligent agent is assigned a subtask of determining the best multicast routing from the source node to one or more of the destination nodes based on the link state information.
	\item The knowledge plane stores the multicast routes.
	\item Before the next network traffic arrives, the control plane distributes flow tables to wireless access nodes in the data plane. Finally, the data plane completes traffic forwarding.
\end{enumerate}

\subsubsection{Data plane}
The data plane is composed of wireless access nodes (APs) and stations (STAs), which perform a set of basic tasks, such as AP-to-controller mapping, packet routing, and site migration tasks, based on instructions issued by the controller. Each AP in the data plane operates without knowledge of the other APs in the wireless network, completely relying on the control plane, application plane, and knowledge plane to perform related operations. It periodically interacts with the controller and transmits wireless network status information to the control plane.

\subsubsection{Control plane}
The control plane contains a centralized controller, which controls and manages the data plane through its southbound interface and constructs a global view of the network in accordance with the network flow and state information from the wireless APs in order to further realize the scheduling of the network resources. The controller also has a northbound interface through which it can interact with the knowledge plane, which facilitates the distribution and deployment of knowledge plane policies. It includes three modules: a network topology discovery module, a link information detection module, and a flow table installation module.

\begin{itemize}
	\item \textbf{Network topology discovery module:} Topology discovery is performed through the OpenFlow Discovery Protocol (OFDP), in which the controller periodically sends Link Layer Discovery Protocol (LLDP) request packets to the data plane to obtain the current network topology and collect information about the connections between network devices. Specifically, the controller sends a Features-Request message to a wireless AP to request its configuration information. Upon receiving the message, the AP encapsulates its port information, MAC address information, and datapath ID information into a Features-Reply packet, which is sent to the controller. The controller then parses this packet to establish a connection with the AP. Based on the collected information, the network topology discovery module establishes associations among the network devices and infers the network topology. It also stores the collected network device status and configuration information for future use.
	
	\item \textbf{Link information detection module:} This module periodically sends status request packets to the devices in the data plane. When a device receives a status request message, it encapsulates its current status information (such as the sizes of sent and received data streams, the number of dropped packets, the congestion status, and the distances to APs) into a data packet and sends it to the controller. The link information detection module then receives the reply messages and parses out the original data containing the network status information from the message packets. The parsed data are also provided to the application plane for processing.
	
	\item \textbf{Flow table installation module:} First, the controller receives the optimal multicast routes selected by the knowledge plane through its northbound interface. Then, before the next data streams arrive, the controller uses its southbound interface to install the flow table entries and send them to the wireless APs. Finally, the data plane forwards the traffic based on the installed flow table entries.	
\end{itemize}

\subsubsection{Application plane}
The application plane primarily handles the data processing logic between the control plane and the knowledge plane. It mainly processes the raw network status data collected from the data plane by the control plane into the network traffic matrix that the knowledge plane requires.

The raw network status data include the numbers of transmitted packets $ tx_p $ and received packets $ rx_p $ for each port, the numbers of transmitted bytes $ tx_b $ and received bytes $ rx_b $, the numbers of dropped packets $ tx_{drop} $ and $ rx_{drop} $, he numbers of erroneous packets $ tx_{err} $ and $ rx_{err} $, and the duration of time $ t_{dur} $ for which the port sends data. Using the collected port status data, the application plane calculates the residual bandwidth $ bw_{ij} $, used bandwidth $ used\_bw_{ij} $, packet loss rate $ loss_{ij} $, packet error rate $ errors_{ij} $, and packet drop rate $ drops_{ij} $ between node $ i $ and node $ j $. The residual bandwidth $ bw_{ij} $ can be calculated by subtracting the used bandwidth $ used\_bw_{ij} $ from the maximum bandwidth $ bw_{max} $ of the link, where the used bandwidth can be derived from $ tx_b $, $ rx_b $, and $ t_{dur} $. The calculation is shown in Equation \eqref{eq12} and Equation \eqref{eq13}.

\begin{equation}
used\_b{w_{ij}} = \frac{{\left| {\left( {t{x_{bi}} + r{x_{bi}}} \right) - \left( {t{x_{bj}} + r{x_{bj}}} \right)} \right|}}{{{t_{durj}} - {t_{duri}}}}
\label{eq12}
\end{equation}

\begin{equation}
b{w_{ij}} = b{w_{max}} - used\_b{w_{ij}}
\label{eq13}
\end{equation}
where $ tx_{bi} $ and $ tx_{bj} $ represent the numbers of bytes transmitted by node $ i $ and node $ j $, respectively; $ rx_{bi} $ and $ rx_{bj} $ represent the number of bytes received by node $ i $ and node $ j $, respectively; and $ t_{duri} $ and $ t_{durj} $ represent the durations of data transmission by the ports of node $ i $ and node $ j $, respectively.

The packet loss rate $ loss_{ij} $ of the link is then calculated from the number of sent packets $ tx_p $ and the number of received packets $ rx_p $, as shown in Equation (14).

\begin{equation}
los{s_{ij}} = \frac{{t{x_{pi}} - r{x_{pj}}}}{{t{x_{pi}}}}
\label{eq14}
\end{equation}
where $ tx_{pi} $ is the number of packets sent by node $ i $ and $ rx_{pj} $ is the number of packets received by node $ j $.

The drop rate $ drops_{ij} $ and error rate $ errors_{ij} $ are calculated from the numbers of packets dropped when sending $ (tx_{drop}) $ and receiving $ (rx_{drop}) $ and the numbers of packets with errors when sending $ (tx_{err}) $ and receiving $ (rx_{err}) $, respectively, as shown in Equations \eqref{eq15} and \eqref{eq16}.

\begin{equation}
drop{s_{ij}} = \frac{{t{x_{dropi}} + r{x_{dropj}}}}{{t{x_{pi}} + r{x_{pj}}}} \cdot 100\% 
\label{eq15}
\end{equation}
\begin{equation}
error{s_{ij}} = \frac{{t{x_{erri}} + r{x_{errj}}}}{{t{x_{pi}} + r{x_{pj}}}} \cdot 100\%
\label{eq16}
\end{equation}
where $ tx_{dropi} $ and $ tx_{erri} $ represent the numbers of dropped and erroneous packets sent by node $ i $, and $ rx_{dropj} $ and $ rx_{errj} $ represent the numbers of dropped and erroneous packets received by node $ j $.

The raw network status data also include the round-trip delays $ RTT_{rs} $ and $ RTT_{rd} $ between the SDN controller and the source and destination switches, respectively, which are obtained by the SDN controller by means of the LLDP protocol and echo requests with timestamps \cite{b43}. The raw status data also includes the forward transmission delay $ T_{fwd} $ and the reply transmission delay $ T_{reply} $ among the three; in detail, $ T_{fwd} $ is the total transmission delay from the controller to the source switch, from the source switch to the destination switch, and then from the destination switch back to the controller, and $ T_{reply} $ is the reverse reply delay. Using the above data, the correct delay $ delay_{ij} $ between the two switches can be calculated as shown in Equation \eqref{eq17}.

\begin{equation}
dela{y_{ij}} = \frac{{\left( {{T_{fwd}} + {T_{reply}} - RT{T_{rs}} - RT{T_{rd}}} \right)}}{2}
\label{eq17}
\end{equation}

In addition, the distance $ distance_{ij} $ between two wireless APs can be calculated based on their deployment coordinates. Since these parameters have different units of measurement, to avoid one parameter having a disproportionate impact on the others, the max-min normalization method \cite{b44} is used to normalize these parameters, as shown in Equation \eqref{eq18}. In this way, all parameters are scaled to within the range of [0,1].

\begin{equation}
{m_{ij}} = \frac{{{m_{ij}} - \min \left( {TM} \right)}}{{\max \left( {TM} \right) - \min \left( {TM} \right)}}
\label{eq18}
\end{equation}
where $ m_{ij} $ is the normalized value of element of the parameter matrix between node $ i $ and node $ j $.  $ \max(TM) $  and $ \min(TM) $  are the maximum and minimum values in the parameter matrix, respectively.

After the calculation and normalization of these parameters, the traffic matrices required for designing the state spaces of the intelligent agents in the knowledge plane are obtained. This allows the intelligent agents to use more comprehensive network status information for learning.

\subsubsection{Knowledge plane}
The knowledge plane is a core module added to the SDWN architecture, and the multicast routing algorithm proposed in this paper runs on this plane. In the knowledge plane, multicast path calculation is performed through multi-agent cooperation. The knowledge plane obtains the processed traffic matrices from the application plane and converts them into training data for the agents. After training, the reward values obtained by the agents converge, that is, each agent uses these traffic matrices to seek its optimal execution strategy. The construction of the multicast paths is completed through the cooperation of multiple agents, and the multicast paths are then sent to the control plane via the northbound interface.

\begin{figure}[htbp]
	\centering
	\includegraphics[width=0.45\textwidth]{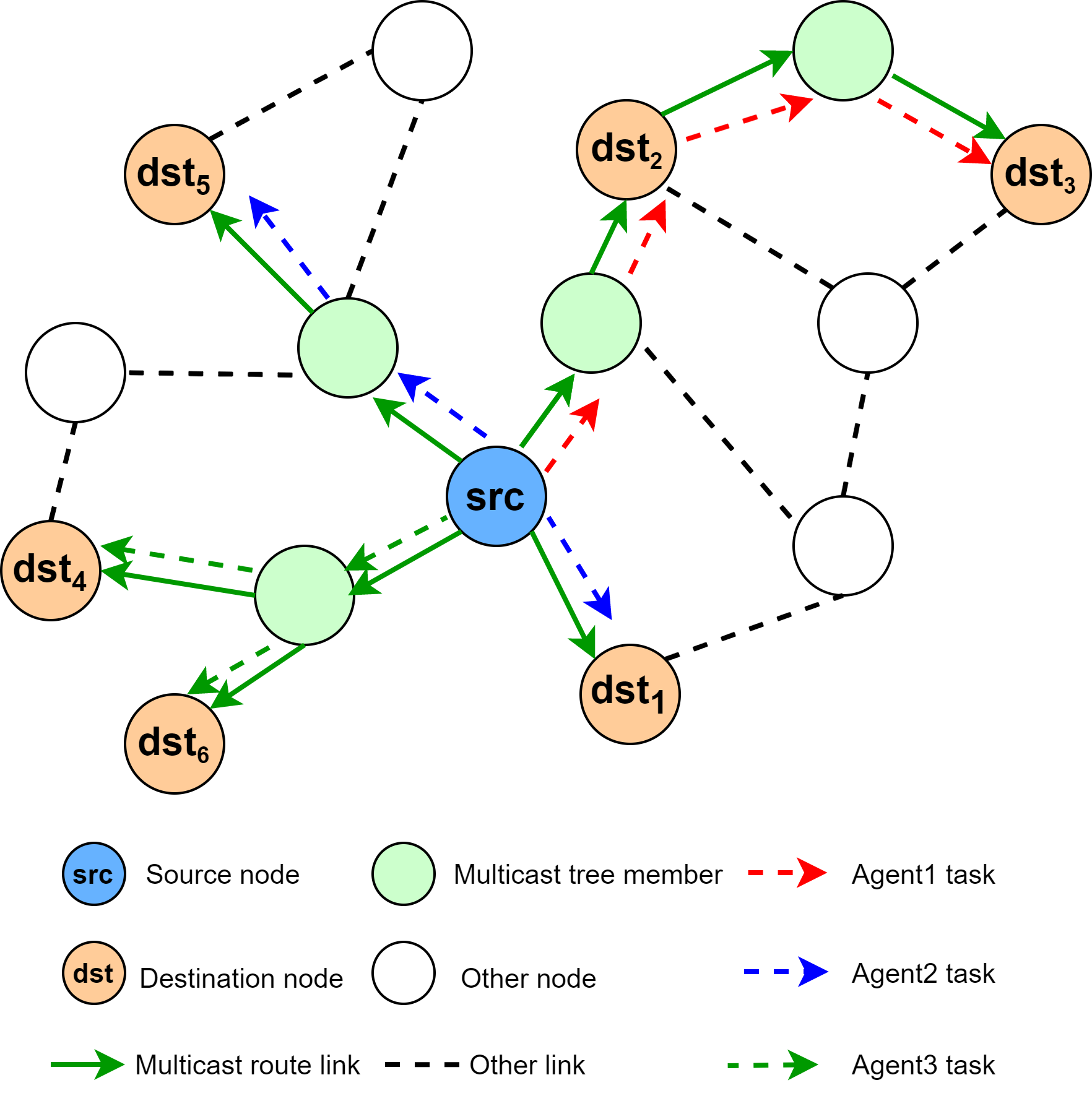}
	\caption{\raggedright Illustrate of multi-agent cooperation for building a multicast tree}
	\label{fig3}
\end{figure}

The scheme for constructing a multicast tree through multi-agent cooperation in the knowledge plane is based on the formal description of the multicast problem given in Section \ref{sec:3.1}. The problem of constructing a multicast tree with the minimum end-to-end costs is decomposed into the construction of multiple unicast paths from the source node to individual destination nodes. The final multicast tree is simply a collection of such unicast paths. To construct such a multicast tree, we abstract the construction process as an MDP \cite{b45}. The state space, action space, and reward function of each intelligent agent are designed using the global network topology and link state information. This is illustrated in Fig. \ref{fig3}.

First, the multicast routes from the source node $ src $ to the multicast destination node set $ DST $, $ DST= \{dst_1,\cdots,dst_6\} $, are decomposed into 6 unicast routes $ \{(src,dst_1),\cdots,(src,dst_6 )\} $. Then, these routes are randomly partitioned among three subtasks $ \{(src,dst_2),(src,\\dst_3)\} $, $ \{(src,dst_1),(src,dst_5)\} $, and $ \{(src,dst_4),(src,dst_6)\} $. These subtasks are randomly assigned to three different agents, i.e., $ agent_1 $, $ agent_2 $, and $ agent_3 $, for completion to obtain the unicast paths $ \{path_1,path_2,\cdots,path_6 \} $ for the corresponding destination nodes. Remove redundant links and ultimately obtain a multicast tree completed by multiple agents working together.

\section{MADRL-MR: AN INTELLIGENT MULTICAST ROUTING ALGORITHM WITH MULTI-AGENT DEEP REINFORCEMENT LEARNING}
\label{sec:MADRL-MR algorithm}
\begin{figure}[htbp]
	\centering
	\includegraphics[width=0.45\textwidth]{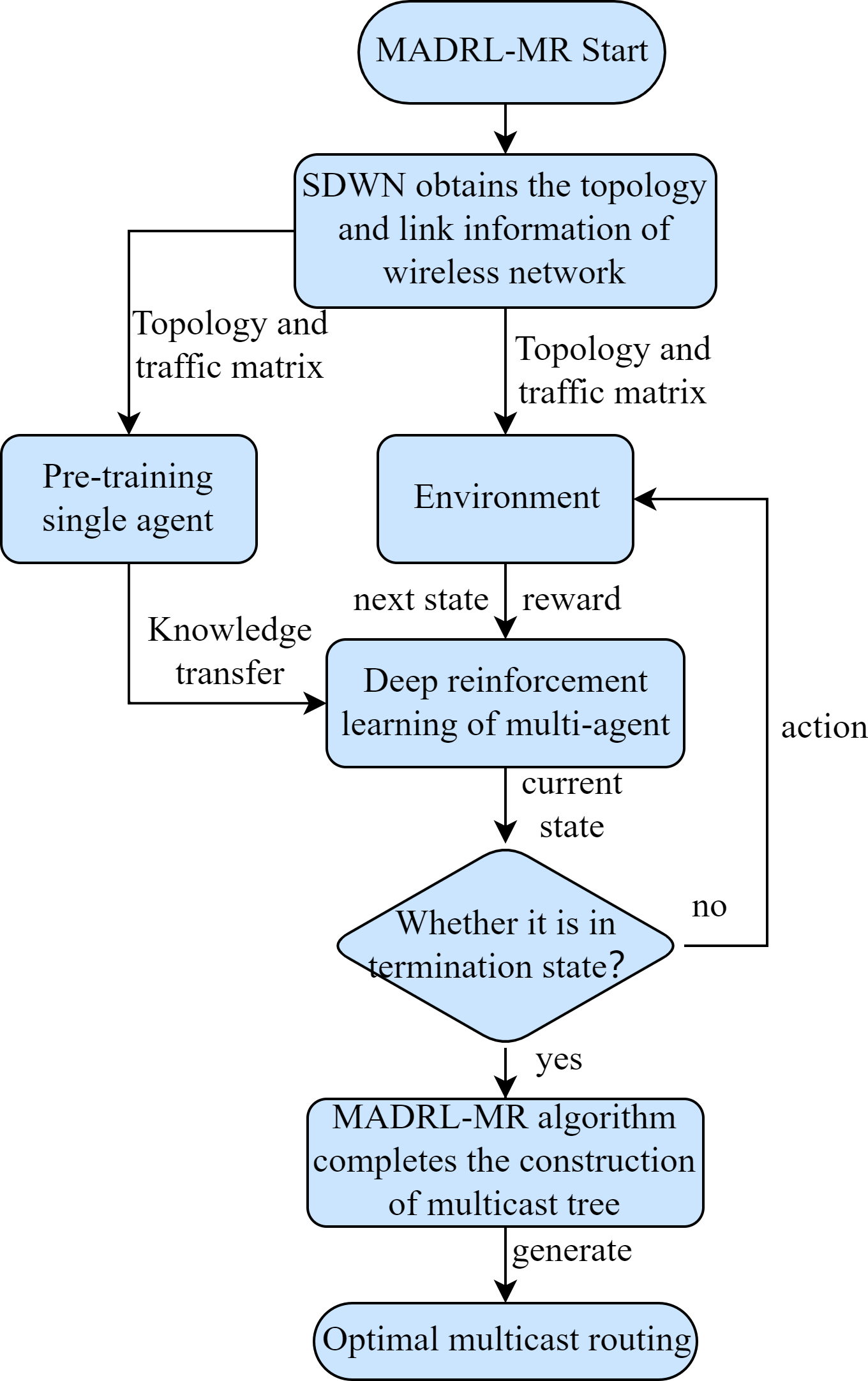}
	\caption{\raggedright Flowchart of the MADRL-MR algorithm}
	\label{fig4}
\end{figure}

The flowchart of the MADRL-MR algorithm is shown in Fig. \ref{fig4}. First, SDWN technology is used to obtain the topology and link state information of the wireless network, and form an environment with which the intelligent agents can interact is formed based on the generated network topology and traffic matrices. A set of pretrained unicast routing agent weights for all nodes is also generated. Each agent in MADRL-MR loads these pretrained weights and interacts with the established environment to obtain the current state. If the state is not a terminal state, then the agent generates an action and interacts with the environment again to obtain the next state and reward. This process is repeated until all agents reach the goal state, that is, the optimal multicast routing paths are generated to construct the multicast tree from the source to all destination nodes.

\begin{figure}[htbp]
	\centering
	\includegraphics[width=0.45\textwidth]{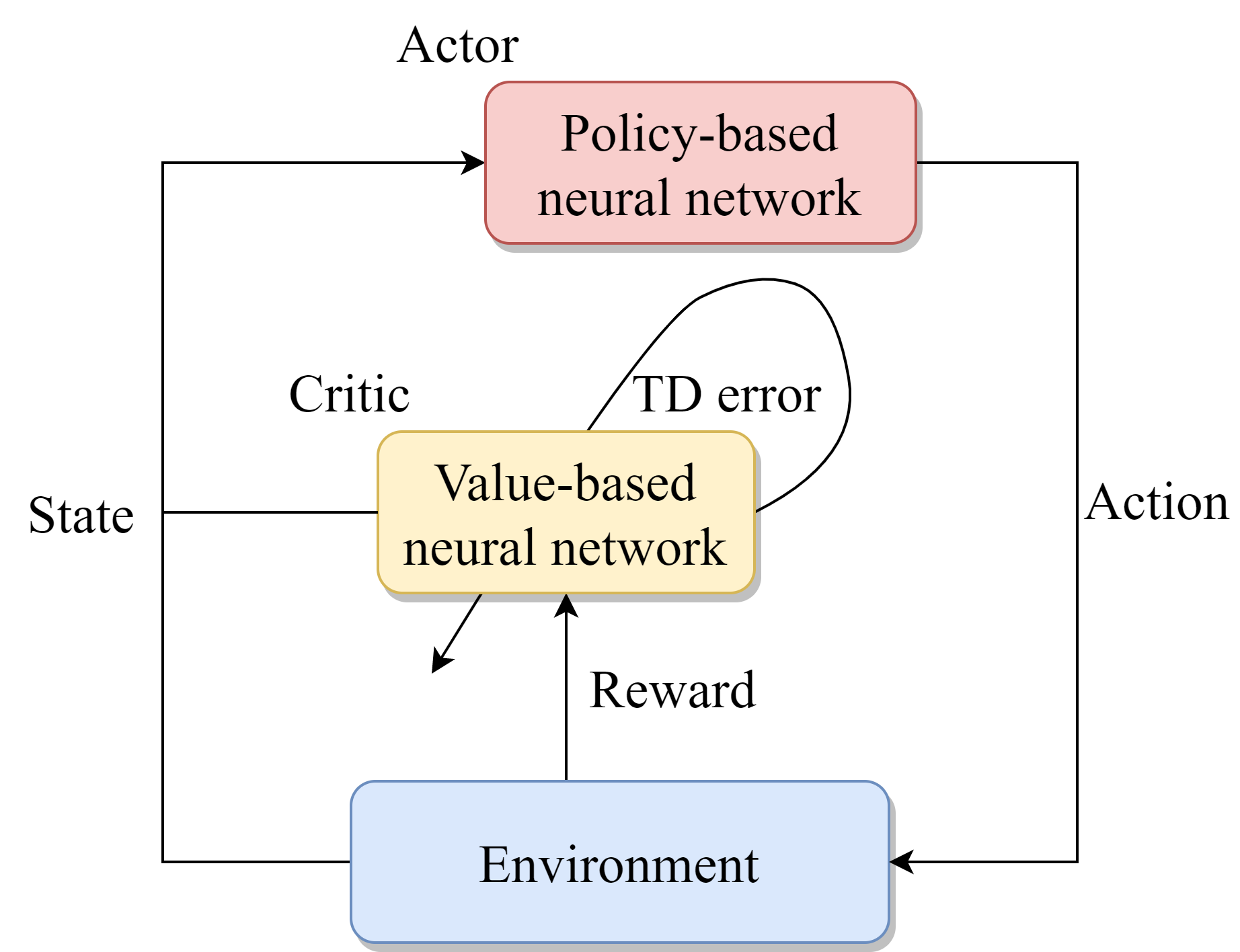}
	\caption{\raggedright Structure of A2C}
	\label{fig5}
\end{figure}

Each agent in MADRL-MR uses the Advantage Actor–Critic (A2C) algorithm \cite{b46} as its core architecture, as shown in Fig. \ref{fig5}. The actor is the policy-based neural network, and the critic is the value-based neural network. A2C is a reinforcement learning method that combines policy gradients and temporal difference learning. It uses an on-policy learning approach to interactively learn from the environment. The learning process involves a series of actions taken to proceed from the source node to the destination node. The selection of each action (i.e., from the current state to the next state) generates a probability distribution for the selection of all possible actions, yielding a policy $ \pi $. The initial actor interacts with the environment to collect data, and based on these data, the value function is estimated using the temporal difference (TD) method. The critic judges the goodness of the selected action in the current state and then updates the policy $ \pi $ based on the value function. Finally, a policy will be trained to select the action with the highest reward value in each state.

The following description of the proposed algorithm starts with the design of the state space, action space, and reward function for each agent and a detailed analysis of the policy gradient update process of A2C. Finally, the designed multiagent training method is introduced, and how transfer learning is used to accelerate the convergence of the designed multiagent algorithm is described.

\subsection{Designed Reinforcement Learning Agents}
\label{sec:4.1}
The MADRL-MR algorithm represents an extension from single-agent to multi-agent reinforcement learning. The design of each reinforcement learning agent includes its state space, action space, reward function, and internal structure, which are identical for all agents in the multi-agent system. Here, we introduce the design of the reinforcement learning agents.

\subsubsection{State space}
\begin{figure}[htbp]
	\centering
	\includegraphics[width=0.45\textwidth]{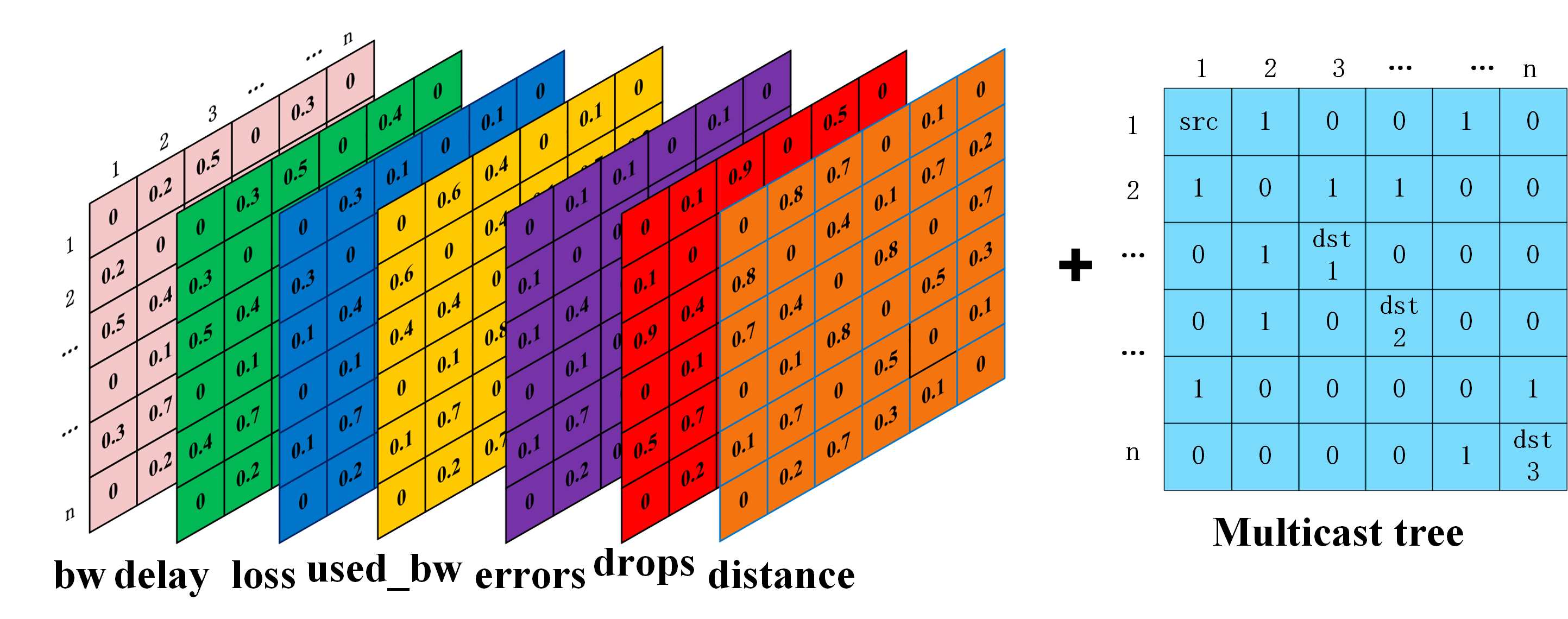}
	\caption{\raggedright The designed state matrix of each agent}
	\label{fig6}
\end{figure}

The state space is a description of the environment and the agent, and the agent can obtain the current state by observing the environment. In the reinforcement learning problem of interest here, the environment consists of the data plane, and the agent's state space is composed of the link state information and the constructed paths from the current source node to all destination nodes in the data plane. We transform this information into a multi-channel matrix $ G_T $, which consists of seven matrices corresponding to different channels, as shown in Fig. \ref{fig6}.

As illustrated in Fig.\ref{fig6}, we transform each of the seven types of data collected from the data plane, namely, $ bw $, $ delay $, $ loss $, $ used\_bw $, $ errors $, $ drops $ and $ distance $, into an adjacency matrix of a weighted undirected graph, resulting in seven traffic matrices with different weights. Here, $ n $ is the number of nodes in the network topology. We also express the constructed multicast tree as a symmetric matrix $ M_{tree} $, where $ src $ is the source node and $ \{dst_1,dst_2,dst_3 \} $ are the destination nodes of the tree. If the matrix element corresponding the edge between node $ i $ and node $ j $ is set to 1, this indicates that this edge is present in the multicast tree.

The set of all possible changes to $ G_T $ is the state space $ \mathcal{S} $. Each state corresponds to a multi-channel matrix. A state transition corresponds to adding a new link to a path, i.e., adding a new link between two nodes. Based on the changes in $ G_T $, the current state $ s_t $ is transformed into the next state $ s_{t+1} $. When paths to all destination nodes have been found (i.e., when the multicast tree has been constructed), $ s_{t+1} $ is set to the terminal state, i.e., $ s_{t+1}=None $.

\subsubsection{Action space}
The action space is the set of actions that an agent can take based on its observation of the current state. In this article, the wireless AP nodes (i.e., the possible next hops) in the data plane are regarded as actions, i.e., $ \mathcal{A}=\{a_1, a_2,\cdots,a_i,\cdots, a_n \}=\{N_1,N_2,\cdots,N_i,\cdots,N_k\} $, where $ N_k $ represents AP node $ i $, for $ i=1,2,\cdots,n $, and $ a_i $ corresponds to $ N_i $. Taking a certain action means adding that action (wireless AP node) to the path from the source node to the destination node. For each state $  s_i \in \mathcal{S}  $ in the state space, the agent can take any action $ a \in \mathcal{A} $, and the execution of action $ a $ will result in a change in the state. In theory, all nodes can be considered as actions, but not all actions can be executed. To meet the requirements of the input tensor for a neural network, some invalid actions will be generated when selecting actions. Suppose that the agent selects a node as an action $ a_t $ that is not adjacent to any node in the current state $ s_t $. If the resulting state $ s_{t+1} $ generated by interacting with the environment does not advance the construction of the path, then this node cannot be added to the multicast tree. Therefore, the agent's valid actions correspond to the set of adjacent nodes in the current state, i.e., the degree $ dr(N_i) $ of the nodes in that state.

\subsubsection{Reward function}
The reward function guides the agent to choose the maximum reward in order to obtain the optimal policy. It measures the value of a certain action taken by the agent in a certain state, thus helping the agent evolve toward an optimal policy. The optimization objective of maximizing the remaining bandwidth and minimizing the delay, packet loss rate, used bandwidth, packet error rate, packet drop rate, and distance between APs is communicated to the agent through the reward function. At each time step, the agent selects an action $ a_t $ in the current state $ s_t $ based on its policy $ \pi $, and the environment responds to this action, resulting in a state transition to $ s_{t+1} $ and the agent receiving a reward value $ r_{t+1} $. When an agent interacts with the environment, it may select either valid or invalid actions. A valid action can lead to a process state, a normal state change, or a terminal state. Thus, there are four possible outcomes that can arise from the interaction between an agent and the environment: a process state ($ PART $), an invalid action ($ HELL $), a loop ($ LOOP $), and a terminal state ($ END $).

\begin{itemize}
	\item 	Process state $ PART $: When the agent executes a valid action and adds a new node to the path, the state transitions to a non-terminal process state, and the agent updates its policy to continue exploring and learning. The reward value is $ R_{part} $ (Equation \eqref{eq19}). To adapt to the dynamic network state and enable the agent to select the optimal combination of actions, we calculate the reward value based on the remaining bandwidth $ bw_{ij} $, the delay $ delay_{ij} $, the packet loss rate $ loss_{ij} $, the used bandwidth $ used\_bw_{ij} $,the packet error rate $ errors_{ij} $, the packet drop rate $ drops_{ij} $, and the distance $ distance_{ij} $ between node $ i $ and node $ j $ on the network link. The weighting factors of these parameters are denoted by $ \beta_l \in [0,1]$, for $ l=1,2,\cdots,7 $. These parameters are all normalized to [0,1] using the max-min method, for which the specific calculation is shown in \eqref{eq18}.
	
	\begin{equation}
	\begin{array}{c}
	{R_{part}} = {\beta _1}b{w_{ij}} + {\beta _2}\left( {1 - dela{y_{ij}}} \right) + {\beta _3}\left( {1 - los{s_{ij}}} \right)\\
	\\
	+ {\beta _4}\left( {1 - used\_b{w_{ij}}} \right) + {\beta _5}\left( {1 - error{s_{ij}}} \right)\\
	\\
	+ {\beta _6}\left( {1 - drop{s_{ij}}} \right) + {\beta _7}\left( {1 - distanc{e_{ij}}} \right)
	\end{array}
	\label{eq19}
	\end{equation}
	
	\item 	Invalid action $ HELL $: When the agent selects an invalid action, that is, selects a non-neighbor node or a node that is already in the multicast tree, the action will not be executed, and the state will remain unchanged. To discourage the agent from selecting invalid actions, a fixed penalty value of $ R_{hell}=C_1 $ is given in this case.
	
	\item 	Loop state $ LOOP $: In addition to process states, there is also a certain probability that executing a valid action will cause the path to form a loop. In this case, although the chosen action is a neighbor node of the current state, once the action is executed, the agent will be trapped in a loop and unable to explore further to find the optimal path. Therefore, the state is rolled back to $ s_t $, and a fixed penalty value of $ R_{loop}=C_2 $ is given.
	
	\item 	Terminal state $ END $: When an action is executed and the paths from the source node to all destination nodes have been found, that is, the multicast tree has been constructed, the state is set to a terminal state, that is,  $ s_{t+1}=None $. In this state, the reward function uses the network link status on the paths of the entire multicast tree to calculate the reward value, as shown in Equation \eqref{eq20}.
	
	\begin{equation}
	\begin{array}{c}
	{R_{end}} = {\beta _1}b{w_{tree}} + {\beta _2}\left( {1 - dela{y_{tree}}} \right)\\
	\\
	+ {\beta _3}\left( {1 - los{s_{tree}}} \right) + {\beta _4}\left( {1 - used\_b{w_{tree}}} \right)\\
	\\
	+ {\beta _5}\left( {1 - error{s_{tree}}} \right) + {\beta _6}\left( {1 - drop{s_{tree}}} \right)\\
	\\
	+ {\beta _7}\left( {1 - distanc{e_{tree}}} \right)
	\end{array}
	\label{eq20}
	\end{equation}
	where $ bw_{tree} $, $ delay_{tree} $, $ loss_{tree} $, $ used\_bw_{tree} $, $ errors_{tree} $, $ drops_{tree} $, $ distance_{tree} $ represents the remaining bandwidth, delay, packet loss rate, used bandwidth, packet error rate, packet loss rate, and the distance for the entire multicast tree, respectively.
	
\end{itemize}

\subsubsection{A2C network parameter update}
The actor network serves as a policy function $ \pi_\theta(a|s) $, where the policy is parameterized as a neural network with $ \theta $ representing its parameters. Given the current state, the network outputs the next action to be taken. The training objective of the network is to maximize the expected cumulative reward. The policy gradient for this network is given by Equation \eqref{eq21}.

\begin{equation}
\begin{array}{c}
\nabla J\left( \theta  \right) = \frac{1}{N}\sum\limits_{n = 1}^N {\sum\limits_{t = 1}^{{T_n}} {\left( {{Q^{{\pi _\theta }}}\left( {s_t^n,a_t^n} \right)} \right.} } \\
\\
\left. { - {V^{{\pi _\theta }}}\left( {s_t^n} \right)} \right)\nabla \log {\pi _\theta }\left( {a_t^n|s_t^n} \right)
\end{array}
\label{eq21}
\end{equation}
where $ Q^{\pi_\theta}(s_t^n,a_t^n ) $ is the expected cumulative return and $ V^{\pi_\theta} (s_t^n ) $ is the expected value of $ Q^{\pi_\theta} (s_t^n,a_t^n ) $ resulting from performing all actions in the state $ s_t^n $.

In actor–critic (AC) algorithms, high variance can occur because not all actions with positive rewards in a single action trajectory may necessarily be optimal; they could instead be suboptimal. To address this issue, A2C introduces a baseline, represented by $ V^\pi(s) $, which is subtracted from the original reward value to calculate the advantage function, as shown in Equation \eqref{eq22}.

\begin{equation}
{A^\theta }\left( {s_t^n,a_t^n} \right) = {Q^{{\pi _\theta }}}\left( {s_t^n,a_t^n} \right) - {V^{{\pi _\theta }}}\left( {s_t^n} \right)
\label{eq22}
\end{equation}

From \eqref{eq22}, it can be seen that two types of estimates are needed: the action–value function estimate $ Q^{\pi_\theta} (s_t^n,a_t^n ) $ and the state–value function estimate $ V^{\pi_\theta} (s_t^n ) $. The expected Q-value calculation method is based on the equation $ Q^{\pi_\theta} (s_t^n,a_t^n )=E[r_t+\gamma V(s_{t+1})] $, where $ \gamma $ is a discount factor satisfying $ \gamma \in [0,1] $. Since the next state $ s_{t+1} $ is updated in the next time step after an action is taken and the reward $ r_t $ is obtained in the current time step, the Q-value is calculated as the expected value of the reward plus the discounted value of the next state by introducing the TD error method, as shown in Equation \eqref{eq23}.

\begin{equation}
\begin{array}{l}
{Q^{{\pi _\theta }}}\left( {s_t^n,a_t^n} \right) = r_t^n + \gamma {V^{{\pi _\theta }}}\left( {s_{t + 1}^n} \right)\\
\\
T{D_{error}} = r + \gamma V\left( {{s_{t + 1}}} \right) - V\left( {{s_t}} \right)\\
\\
{A^\theta }\left( {s_t^n,a_t^n} \right) = r_t^n + \gamma {V^{{\pi _\theta }}}\left( {s_{t + 1}^n} \right) - {V^{{\pi _\theta }}}\left( {s_t^n} \right)
\end{array}
\label{eq23}
\end{equation}

The critic network calculates the value function $ V^\pi (s) $, which represents the future payoff that the agent can expect from state $ s $, and estimates the value function of the current policy from this expected payoff, that is, it evaluates the goodness of the actor network. With the help of the value function, an AC algorithm can perform a single-step parameter update without waiting until the end of the round. The value function is calculated as shown in Equation \eqref{eq24}.

\begin{equation}
{V^\pi }\left( s \right) = {E_\pi }\left\{ {{r_t}|{s_t} = s} \right\}
\label{eq24}
\end{equation}

The critic network parameters $ \omega $ are updated using the mean squared error loss function through backward propagation of the gradient, where the mean squared error loss function is shown in Equation \eqref{eq25}.

\begin{equation}
MSE = {\sum {\left( {r + \gamma {V^\pi }\left( {{s_{t + 1}}} \right) - {V^\pi }\left( {{s_t},\omega } \right)} \right)} ^2}
\label{eq25}
\end{equation}

According to the policy gradient formula analyzed above, Equation \eqref{eq26} for updating the parameters of the policy function of the actor network is obtained by combining the comparative advantage function of A2C with the TD method, where $ \alpha_1 $ is the learning rate.

\begin{equation}
\begin{array}{r}
\theta  = \theta  + {\alpha _1}\frac{1}{N}\sum\limits_{n = 1}^N {\sum\limits_{t = 1}^{{T_n}} {\left( {{Q^{{\pi _\theta }}}\left( {s_t^n,a_t^n} \right)} \right.} } \\
\\
\left. { - {V^{{\pi _\theta }}}\left( {s_t^n} \right)} \right)\nabla \log {\pi _\theta }\left( {a_t^n|s_t^n} \right)
\end{array}
\label{eq26}
\end{equation}

\subsubsection{Multi-agent training strategy for MADRL-MR design}
\begin{figure}[htbp]
	\centering
	\includegraphics[width=0.45\textwidth]{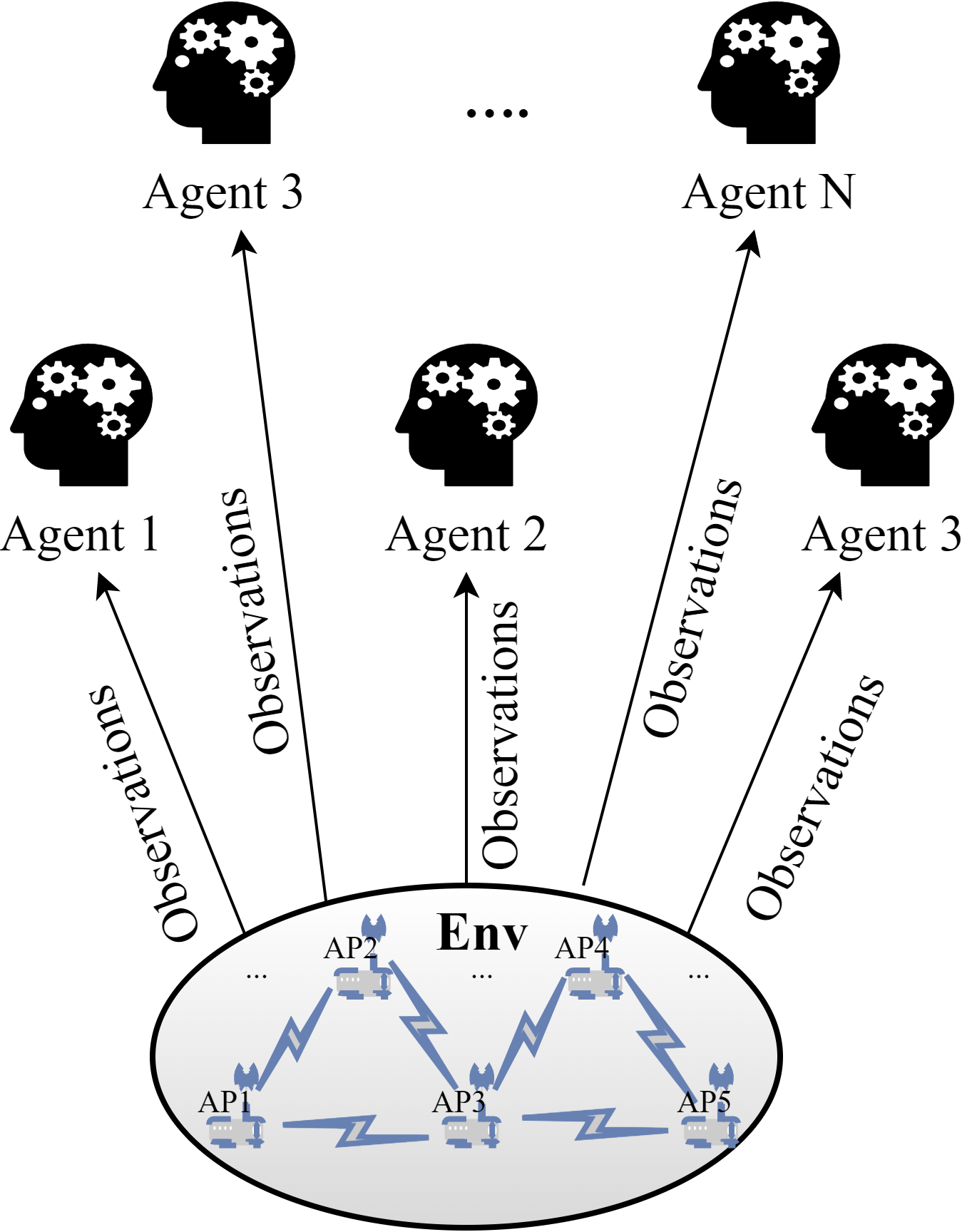}
	\caption{\raggedright Fully decentralized training}
	\label{fig7}
\end{figure}

Four main challenges are encountered in multi-agent reinforcement learning: non-stationarity of training, scalability, partial observability, and privacy and security. Based on the above challenges, multi-agent training methods can be divided into three main types: fully decentralized (IL) training, fully centralized training, and centralized training and decentralized execution (CTDE) \cite{b47}. Although fully centralized training alleviates the issues of partial observability and non-stationarity, it is not feasible for large-scale and real-time systems. Moreover, since the CTDE method relies on a centralized control unit that collects training information from each agent, it also has difficulty scaling to environments with large numbers of agents. With an increasing number of agents, a centralized critic network will suffer from increasingly high variance, and the value function will have difficulty converging. Therefore, a fully decentralized training method is adopted in this paper, as shown in Fig. \ref{fig7}.

This method is a direct extension from the single-agent scenario to the multiple-agent scenario, in which each agent independently optimizes its policy without considering non-stationarity issues. To address the convergence challenges of this method for the agents, we adopt the strategy of transfer reinforcement learning. In practice, the IL method has achieved satisfactory results for several resource allocation and control problems in wireless communication networks \cite{b48,b49,b50}.

\subsubsection{Reinforcement Learning with Transfer Learning Mechanisms}
To accelerate the training speed of the multi-agent system and address the issues of instability and difficulty in convergence during the training phase, this paper applies transfer learning in combination with reinforcement learning. Transfer learning (TL) allows knowledge acquired from experts or other processes to be transferred to the current task, which accelerates learning. The applications of transfer learning in reinforcement learning can be divided into the following three main categories depending on the transfer setting \cite{b51}: $ \textcircled{1} $ fixed-domain transfer from a single source task to a target task, $ \textcircled{2} $ fixed-domain migration across multiple source tasks to target tasks, and $ \textcircled{3} $ transfer between source and target tasks in different domains.

This paper adopts the second approach, which involves fixed-domain transfer across multiple source tasks to a target task in the same task domain. Specifically, a pretraining process is conducted to obtain the initial weights of an intelligent agent for single-broadcast routing that covers all source nodes and destination nodes with the same state space, action space, and reward function. In MADRL-MR, each intelligent agent loads these initial weights before learning, in a process called knowledge transfer, to reduce ineffective exploration at the beginning of training. Then, during the training process, the algorithm parameters are adjusted based on the different tasks of the multiple intelligent agents to accelerate their convergence. This approach aims to enable stable coordination among multiple intelligent agents and make them more adaptable to dynamic changes in the network.

\subsection{Design of the MADRL-MR algorithm}
\label{sec:4.2}
In the MADRL-MR algorithm, the paths between the input source node $ src $ and the destination nodes $ DST $ are first divided among several subtasks such that each agent is assigned different tasks of establishing paths from $ src $ to multiple destination nodes $ dst\in DST $. Second, based on the current network topology $ graph $ (the environment), each agent learns the optimal unicast paths from $ src $ to multiple $ dst $ nodes. Finally, once all agents have completed their tasks, the learned paths from $ src $ to all $ dst $ nodes are combined to obtain the optimal multicast tree from $ src $ to $ DST $ in $ graph $. The detailed implementation of the MADRL-MR algorithm for multicast routing with multiagent deep reinforcement learning is shown in Algorithm \ref{alg:MADRL-MR}.

The algorithm takes as input a network topology $ G(V,E) $, a traffic matrix $ TM $, the source node and destination nodes $ (src,DST) $for multicasting, and hyperparameters for the reinforcement learning algorithm. The output of the algorithm is an optimal multicast tree from the source node $ src $ o the set of destination nodes $ DST $. Lines 1 to 3 initialize the actor network parameters, the critic network parameters, and the experience buffer, respectively, which are discarded after each update. Line 4 uses transfer learning technology to load the pretrained agent weights for all source nodes to all destination nodes obtained before the start of training, i.e., performs knowledge transfer. On line 5, the destination nodes in the multicast group are equally and randomly split among several subtasks according to the number of input agents, and one subtask is assigned to each agent. Lines 8 and 9 initialize the environment to obtain the initial state $ s_t $. Lines 11 to 13 input the state $ s_t $ into the actor network and select an action $ a_t $ based on importance sampling. Then, the selected action is performed to interact with the environment to obtain the reward value $ r_t $ and the next state $ s_{t+1} $, and the experience $ (s_t,a_t,r_t,s_{t+1} ) $ is stored in the experience buffer. Lines 14 to 18 involve learning from the experiences stored in the experience buffer by inputting $ s_t $ and $ s_{t+1} $ into the critic network to obtain $ V^\pi (s_t ) $ and $ V^\pi (s_{t+1} ) $. The TD error, $ TD_{error} $, is then calculated using \eqref{eq23}. Line 19 calculates the mean squared error loss function \eqref{eq25} to be used for the gradient update of the critic network parameters $ \omega $. Line 20 updates the actor network parameters $ \theta $ according to \eqref{eq26} based on $ V^\pi (s_t ) $, $ V^\pi (s_{t+1} ) $ and $ TD_{error} $ . Line 21 clears the experience buffer. Lines 24 to 27 judge whether all agents have found the desired paths from the source node to the destination nodes and obtain the optimal multicast tree by removing redundant links from these paths. Finally, line 28 updates the state $ s_t $ to proceed to the next episode.

\begin{algorithm}[htb]
	\footnotesize
	\caption{MADRL-MR}
	\label{alg:MADRL-MR}
	\begin{algorithmic}[1]
		\Require
		network topology $G(V,E)$, traffic matrix $TM$, source and multicast destination node $(src,DST)$, weight factor $\beta_l,l=1,2,...7,$, actor learning rate $\alpha_1$, critic learning rate $\alpha_2$, reward discount factor $\gamma$, batch-size $k$, update frequency $update_{time}$, number of agents $n$, training episodes $episodes$, pre-training weights of actor and critic $\hat{\theta}$ and $\hat{\omega}$
		
		\Ensure
		optimal multicast tree for $tree(src, DST)$
		
		\State Initialize actor network $\theta$
		\State Initialize critic network $\omega$
		\State Initialize buffer capacity $B$
		\State Load pre-training weights $\theta=\hat{\theta}$, $\omega=\hat{\omega}$
		\State Assign destination nodes to each agent randomly and equally
		
		\For{$episode \leftarrow 1$ to $episodes$}
		\For{$TM$ in Network Information Storage}
		
		\State Reset environment with $(src, DST)$
		\State The agent obtains the initial state $s_t$
		
		\While{True}
		\State Choose an action $a_t$ from $s_t$ by sampling the output action \hspace*{4.5em} probability
		\State Execute action $a_t$ and observe reward $r_t$ and next state $s_{t+1}$
		\State Store {$(s_t, a_t, r_{t}, s_{t+1})$} in $B$
		\If{$len(B)\geq k$}
		\For{$i \leftarrow 1$ to $update_{time}$}
		\State Sample batch $k$ data
		\State Enter $data(s_t)$ and $data(s_{t+1})$ in the critic network 
		\hspace*{7.5em} to get $V^\pi(s_t)$ and $V^\pi(s_{t+1})$
		\State Calculate $TD_{error}$ 
		\Statex \hspace*{9em} $TD_{error} \leftarrow r+\gamma V^\pi(s_{t+1})-V^\pi(s_t)$
		\State Calculate MSE as gradient update of critic network 
		\hspace*{7.5em} parameters $\omega$.
		\Statex \hspace*{9em} $MSE \leftarrow \sum(r+\gamma V^\pi(s_{t+1})-V^\pi(s_t,\omega))^2$
		\State Update actor network parameters $\theta$ according to 
		\hspace*{7.5em} Equation \eqref{eq26}
		\State Empty buffer $B$
		\EndFor
		\EndIf
		\If{done} //The path of all destination nodes 
		\hspace*{12em} has been found
		\State Build a multicast tree
		\State Break
		\EndIf
		\State $s_t \leftarrow s_{t+1}$
		\EndWhile
		\EndFor
		\EndFor
	\end{algorithmic}
\end{algorithm}

\section{EXPERIMENTAL SETUP AND PERFORMANCE EVALUATION}
\label{sec:Experiment}
This section describes the experimental settings used in this study and the corresponding performance evaluation. First, the experimental environment is introduced. Second, the performance metrics for algorithm evaluation are defined. Then, the tuning and setting of the algorithm hyperparameters during the experimental process are described. Finally, the results of comparative experiments are discussed.

\subsection{Experimental environment}
\label{sec:5.1}
For the experimental environment in this study, we used Mininet-WIFI 2.3.1b as the simulation platform for the SDWN network. Mininet-WIFI \cite{b52} is a branch of the Mininet SDN network emulator that extends the functionality of Mininet by adding virtualized wireless APs based on standard Linux wireless drivers and the 80211\_hwsim wireless simulation driver. The SDWN controller used in the experiment is Ryu 4.3.4 \cite{b53}. The experiment was conducted on a server with hardware consisting of a 64-core processor and a GeForce RTX 3090 graphics card and with Ubuntu 18.04.6 as the software environment. The Iperf \cite{b54} tool was used to send User Datagram Protocol (UDP) packets.

\begin{figure}[htbp]
	\centering
	\includegraphics[width=0.45\textwidth]{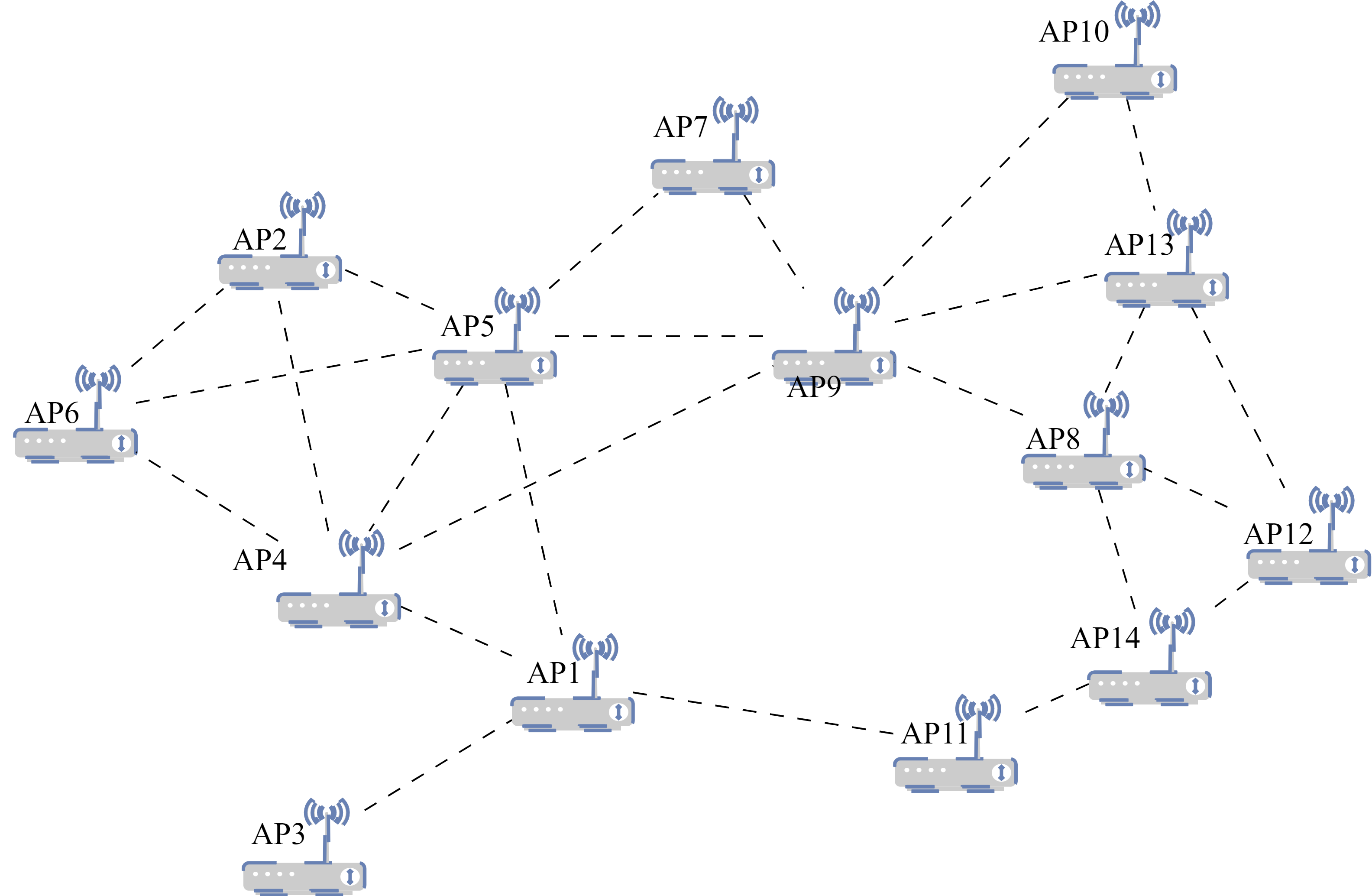}
	\caption{\raggedright Wireless network topology}
	\label{fig8}
\end{figure}

We designed a network topology consisting of 14 wireless nodes to test the performance of MADRL-MR, as shown in Fig. \ref{fig8}. The parameters of the network links were randomly generated following a uniform distribution. The ranges of the random link bandwidth and delay values were 5–40 Mbps and 1–10ms, respectively, while the distances between wireless APs were set within the range of 30–120 m.

\begin{figure}[htbp]
	\centering
	\includegraphics[width=0.45\textwidth]{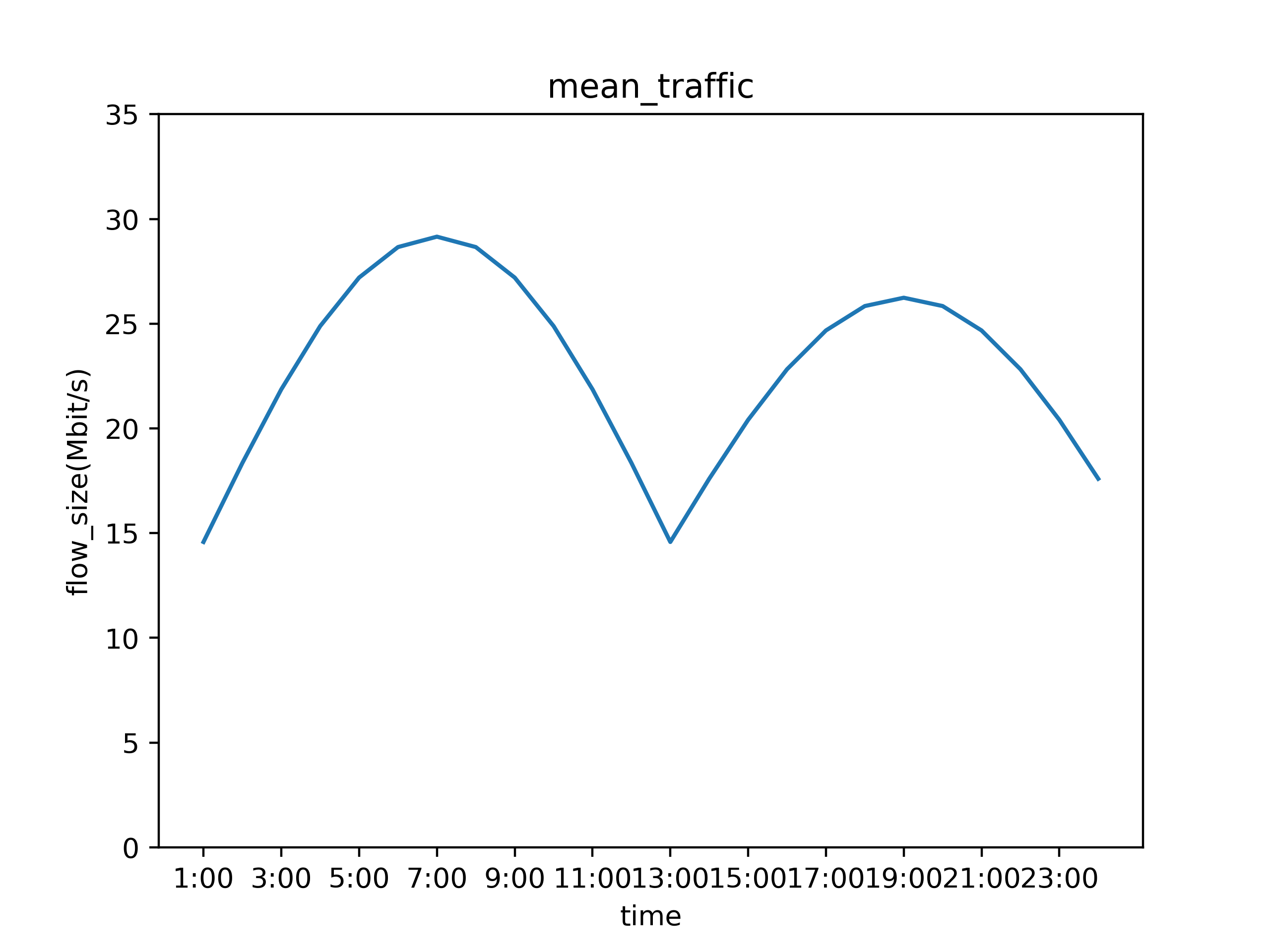}
	\caption{\raggedright Flows sent by Iperf}
	\label{fig9}
\end{figure}

To more accurately simulate a real environment, we used the Iperf traffic generator tool to simulate the network traffic situation 24 hours a day, as shown in Fig. \ref{fig9}. The horizontal axis represents time, and the vertical axis represents the average traffic sent by each node in units of Mbit/s. The traffic distribution conforms to a typical network traffic distribution at different times of day.

\subsection{Performance metrics}
\label{sec:5.2}
As performance indicators, we use the convergence status of the intelligent agents' reward values as well as commonly used performance metrics for routing, such as the instantaneous throughput, delay, and packet loss rate. In addition, we use the remaining bandwidth, tree length, and average distance between wireless APs in the multicast tree as evaluation metrics for our algorithm.

\begin{enumerate}
	\item The calculation of the reward value is described in section \ref{sec:4.1}, specifically, the reader is referred to the formula for the reward function.
	
	\item For the three commonly used evaluation metrics of instantaneous throughput, delay, and packet loss rate, based on the simulated 24-hour network traffic, we use their average values in different time periods to represent the network performance, as shown in Equation \eqref{eq27}.
	
	\begin{equation}
	\begin{array}{c}
	\overline {throughput}  = \frac{{\sum\nolimits_i {\sum\nolimits_j {throughpu{t_{ij}}} } }}{{\Delta t}}\\
	\\
	\overline {delay}  = \frac{{\sum\nolimits_i {\sum\nolimits_j {dela{y_{ij}}} } }}{{\Delta t}}\\
	\\
	\overline {loss}  = \frac{{\sum\nolimits_i {\sum\nolimits_j {los{s_{ij}}} } }}{{\Delta t}}
	\end{array}
	\label{eq27}
	\end{equation}
	where $ avg\_throughput $, $ avg\_delay $, and $ avg\_loss $ represent the average throughput, average delay and average packet loss rate over a time duration $ \Delta t $ time, respectively, and $ throughput_{ij} $ is the throughput from node $ i $ to node $ j $.
	
	\item For the remaining bandwidth, tree length, and average distance between wireless APs in the multicast tree, we use multiple measurements and obtain the average value as the corresponding evaluation metric, as shown in Equation \eqref{eq28}.
	
	\begin{equation}
	\begin{array}{c}
	\overline {b{w_{tree}}}  = average\frac{{\sum\nolimits_n {\sum\nolimits_{ij \in tree} {b{w_{ij}}} } }}{{n \cdot E}}\\
	\\
	\overline {le{n_{tree}}}  = average\frac{{\sum\nolimits_n {le{n_{tree}}} }}{n}\\
	\\
	\overline {dis{t_{tree}}}  = average\frac{{\sum\nolimits_n {\sum\nolimits_{ij \in tree} {distanc{e_{ij}}} } }}{{n \cdot E}}
	\end{array}
	\label{eq28}
	\end{equation}
	where $ \overline {b{w_{tree}}} $ and $ \overline {dis{t_{tree}}} $ represent the average remaining bandwidth per link in the multicast tree and the average distance between wireless APs in the multicast tree, respectively. $ \overline {le{n_{tree}}} $ is the average length of the multicast tree after multiple measurements. $ bw_{ij} $ and $ distance_{ij} $ are the remaining bandwidth and the distance, respectively, from node $ i $ to node $ j $ in the multicast tree. $ n $ is the number measurements performed at a given time. $ E $ is the number of edges in the multicast tree.	
\end{enumerate}

\subsection{Transfer Learning Performance and parameter Settings}
\label{sec:5.3}
\begin{figure}[htbp]
	\centering
	\includegraphics[width=0.45\textwidth]{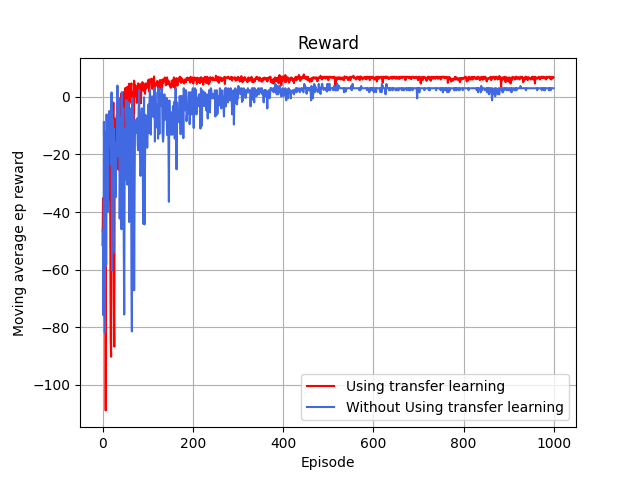}
	\caption{\raggedright Comparison between reward values achieved with and without TL}
	\label{fig10}
\end{figure}

First, the impact of transfer learning on the convergence of the multi-agent SDWN-based intelligent multicast routing algorithm is analyzed, as shown in Fig. \ref{fig10}. The convergence of the reward values is significantly faster with transfer learning than without, and the reward values are also higher with transfer learning. When transfer learning is used in reinforcement learning, a set of pretrained initial weights for connecting all source nodes to all destination nodes in different environments is loaded before each agent starts learning. This is also known as knowledge transfer and endows the agents with some decision-making ability at the beginning of training. In this way, the agents can reach convergence faster than they would if each agent needed to learn from scratch, and it also solves the problem of slower convergence with an increasing number of agents. Therefore, applying transfer learning in multiagent reinforcement learning endows the MADRL-MR intelligent multicast routing algorithm with a stronger learning ability, enables the agents to learn efficient behaviors more quickly, and accelerates the convergence of the reward values. This confirms that transfer learning can improve the performance of the MADRL-MR algorithm.

The most important prerequisite for using deep reinforcement learning to select the optimal multicast routes is the setting of the hyperparameters. We use a multicast group with a complex set of possible paths as a representative example, with node 3 as the source node and multicast destinations of \{6, 7, 8, 9, 11, 13\}. The more complex the set of possible paths to each destination node is, the more choices the agent can explore, and since the multicast tree constructed from the source node to all multicast destination nodes is not unique, testing the effectiveness of the algorithm becomes more challenging with more complex path situations.

The hyperparameter settings for the reward values $ R_{part} $, $ R_{hell} $, $ R_{loop} $, and $ R_{end} $ will affect the convergence speed of the agents. If these hyperparameters are not set appropriately, the agents' reward values may fail to converge, i.e., the agents may be unable to find the optimal strategy.

The first step is to set the penalty values $ R_{hell} $ and $ R_{loop} $. Since the selected actions may be invalid or create loops, this will greatly affect the construction of the multicast tree. Too many invalid actions and loops will affect the convergence speed of the agents and may even lead to nonconvergence. Therefore, the setting of the penalty values is crucial. To reduce the influence of the other two reward values on the setting of the penalty values, we set the weighting factors in the calculation formulas of $ R_{part} $ and $ R_{end} $ to 1. Additionally, since all reward calculation parameters are normalized to [0,1], we initially set both penalty values to -1. We then evaluated the performance achieved under various settings of these two penalty values based on the approximate round when convergence began and the total reward value and adjusted the penalty values multiple times accordingly. The results are shown in Table \ref{tab1}.

\begin{table}
	\centering
	\caption{\textbf{PENALTY SETTING}}
	\setlength{\tabcolsep}{3pt}
	\begin{tabular}{|>{\centering\arraybackslash}p{50pt}|>{\centering\arraybackslash}p{50pt}|>{\centering\arraybackslash}p{50pt}|>{\centering\arraybackslash}p{50pt}|}
		\hline
		$ R_{hell} $& $ R_{loop} $& Episode& Reward \\
		\hline
		-1& -1& 450& -23 \\
		-1& -0.7& 670& -30 \\
		-1& -0.5& 390& 12 \\
		-1& -0.1& 810& -15 \\
		-0.7& -0.5& 250& 22 \\
		-0.5& -0.5& 380& -9 \\
		-0.1& -0.5& 460& -5 \\
		\hline
	\end{tabular}
	\label{tab1}
\end{table}

In multiple rounds of adjustment, we first fixed the value of $ R_{hell} $ to -1 and adjusted the value of $ R_{loop} $. The results showed that with $ R_{loop}=-0.5 $, convergence started at approximately the 390th episode, with a converged value of approximately 12. Second, we fixed $ R_{loop} $ at 0.5 and adjusted $ R_{hell} $. It was found that $ R_{hell}=-0.7 $ gave the best result, with the agents starting to converge at approximately the 250th episode and achieving a convergence value of 22, which was the best result among the tested parameter settings. Therefore, we set the values of $ R_{hell} $ and $ R_{loop} $ to -0.7 and -0.5, respectively.

Next, $ R_{part} $ and $ R_{end} $ need to be set because the reward values for these two cases are calculated based on the traffic matrix of the network links, as shown in \eqref{eq19} and \eqref{eq20}. The design of these two reward values mainly involves setting the weight ratios of the seven network link parameters. A different weighting factors should be set for each network link parameter to represent the influence of that parameter on the construction of the multicast tree. We set the weights for the seven parameters, namely, remaining bandwidth, delay, packet loss rate, used bandwidth, packet error rate, packet drop rate, and distance between APs, to [0.7, 0.3, 0.1, 0.1, 0.1, 0.1, 0.1]. In detail, since our goal is to construct a multicast tree with the main influencing factors being the remaining bandwidth, delay, and packet loss rate, we set the weights of the first three parameters to 0.7, 0.3, and 0.1, respectively. The remaining parameters, namely, the used bandwidth, packet error rate, packet drop rate, and distance between APs, are equally important in building an optimal multicast tree, but compared to the first three parameters, we consider them to be supplementary factors. Therefore, we set the weights of the used bandwidth, packet error rate, packet drop rate, and distance between APs all to 0.1. Thus, the initial weights of all parameters are [0.7, 0.3, 0.1, 0.1, 0.1, 0.1, 0.1]. 

\begin{figure*}[htb]
	\centering
	\begin{minipage}{0.3\textwidth}
		\subfigure[]{
			\centering
			\includegraphics[width=1\textwidth]{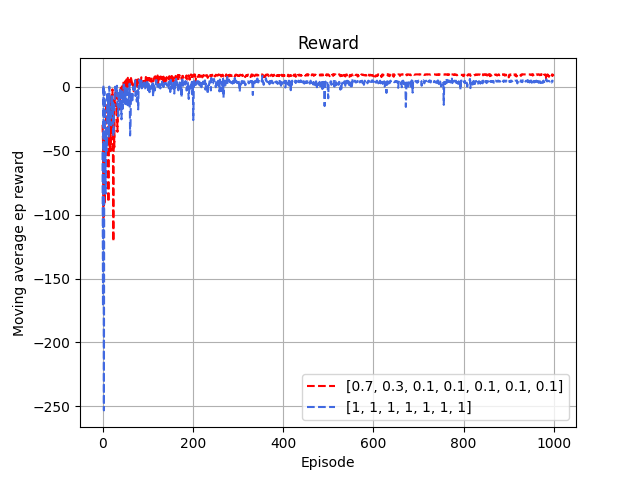}\hspace{-15pt}
			\label{subfig:fig11a}
		}
	\end{minipage}
	\begin{minipage}{0.3\textwidth}
		\subfigure[]{
			\centering
			\includegraphics[width=1\textwidth]{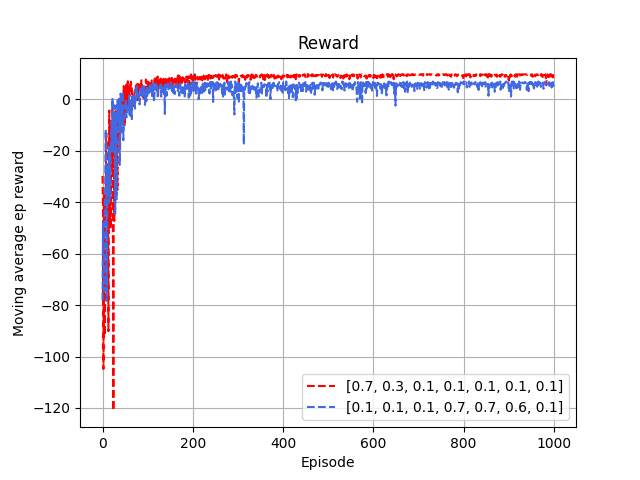}\hspace{-15pt}
			\label{subfig:fig11b}
		}
	\end{minipage}
	\begin{minipage}{0.3\textwidth}
		\subfigure[]{
			\centering
			\includegraphics[width=1\textwidth]{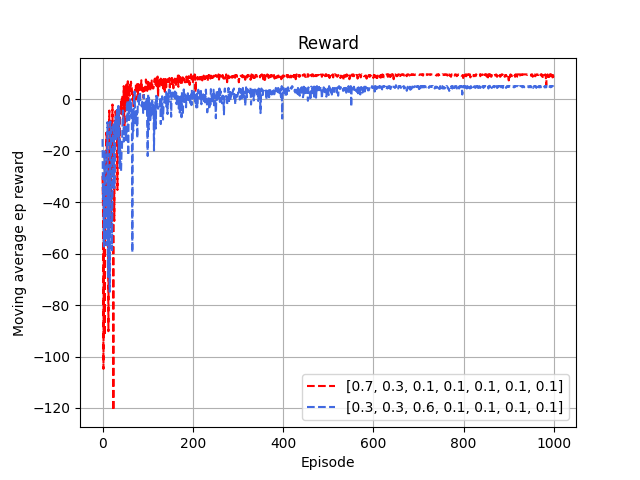}\hspace{-15pt}
			\label{subfig:fig11c}
		}
	\end{minipage}
	\begin{minipage}{0.3\textwidth}
		\subfigure[]{
			\centering
			\includegraphics[width=1\textwidth]{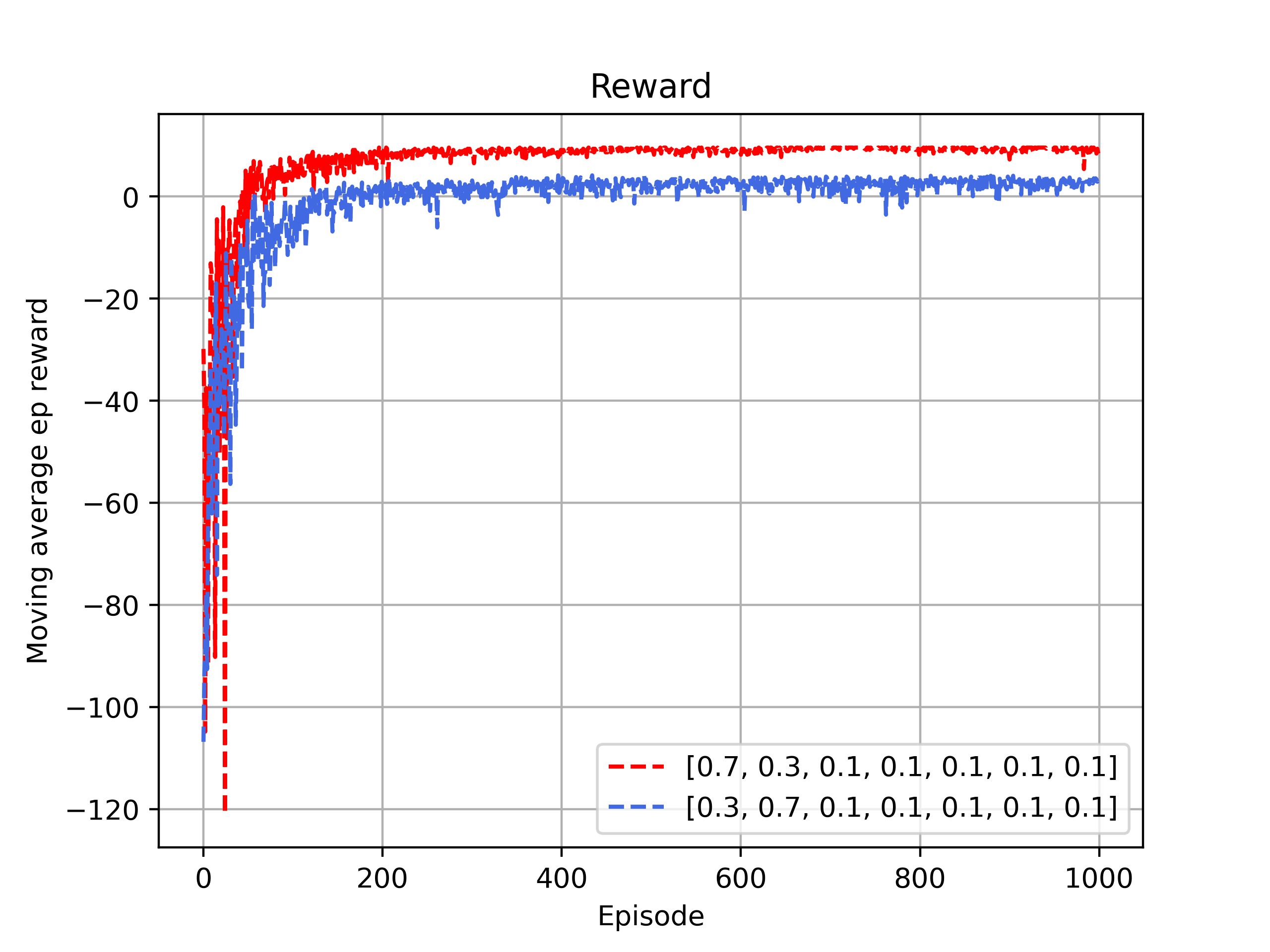}\hspace{-15pt}
			\label{subfig:fig11d}
		}
	\end{minipage}
	\begin{minipage}{0.3\textwidth}
		\subfigure[]{
			\centering
			\includegraphics[width=1\textwidth]{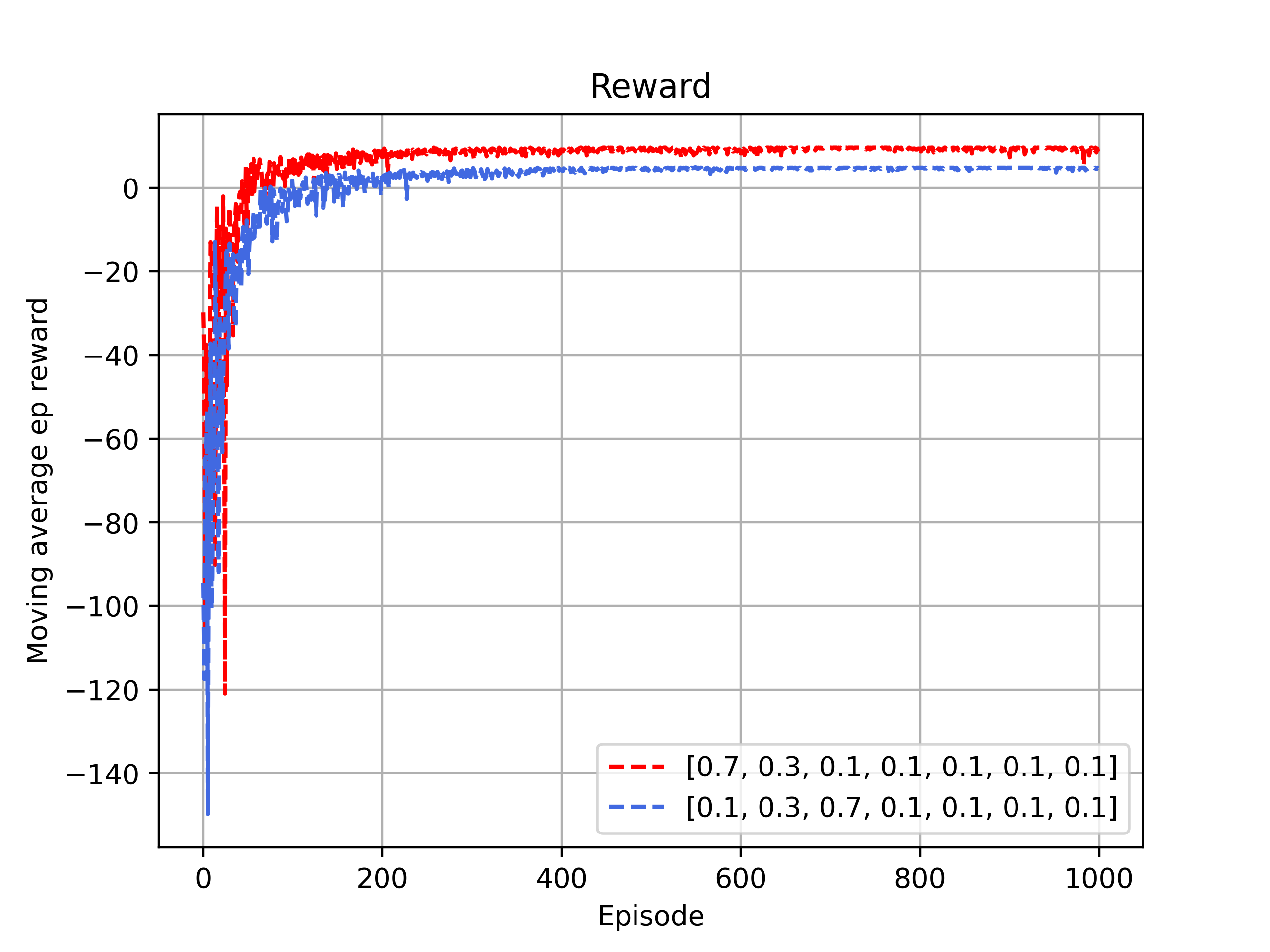}\hspace{-15pt}
			\label{subfig:fig11e}
		}
	\end{minipage}	
	\begin{minipage}{0.3\textwidth}
		\subfigure[]{
			\centering
			\includegraphics[width=1\textwidth]{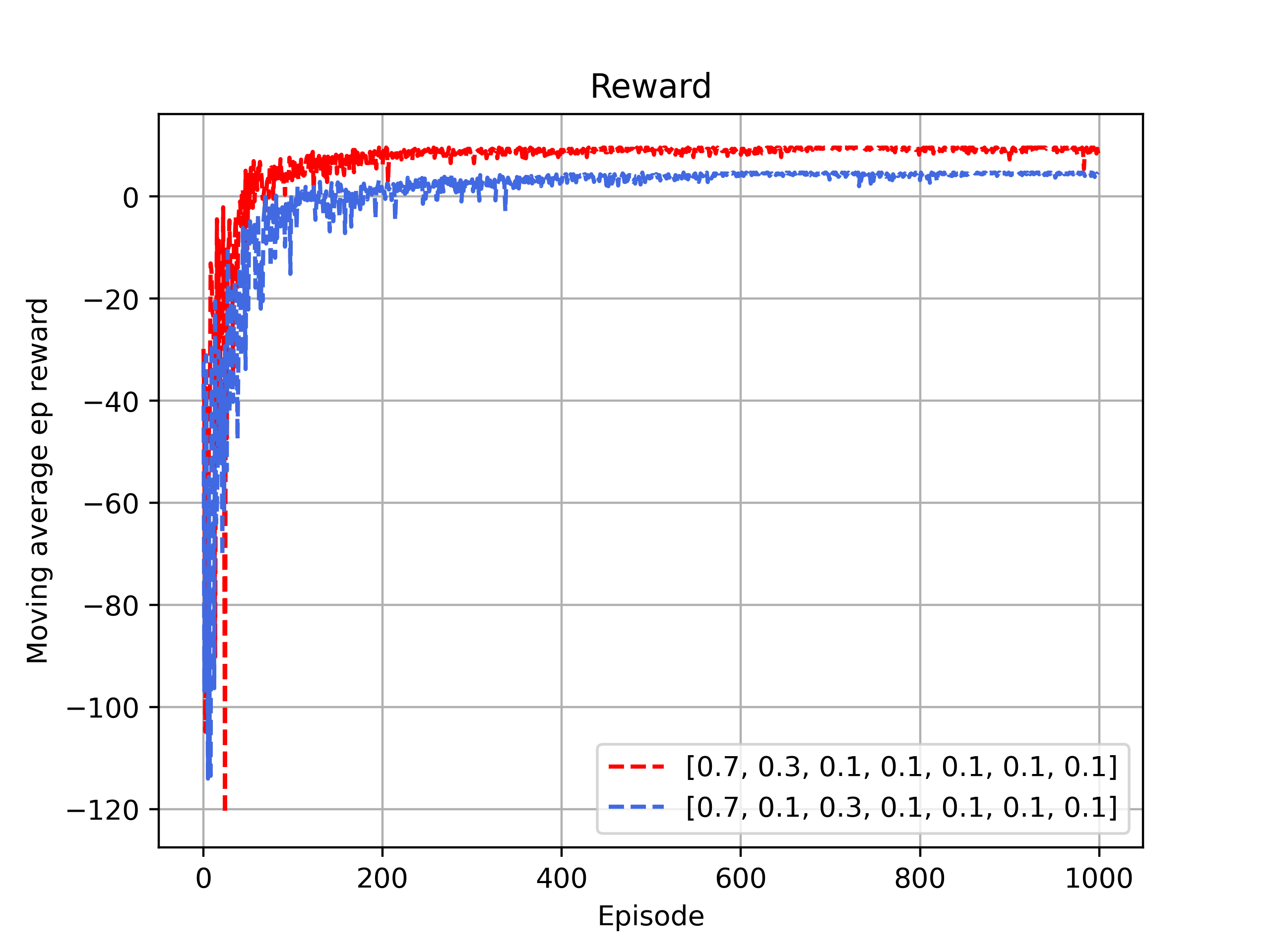}\hspace{-15pt}
			\label{subfig:fig11f}
		}
	\end{minipage}	
	\caption{The weight factor settings of the $ R_{part} $ and $ R_{end} $ reward functions uniformly compare the weights we set [0.7, 0.3, 0.1, 0.1, 0.1, 0.1, 0.1], where (a) is a comparison chart with all weights of 1; (b) is a comparison chart with adjustments to the weights of the last four parameters; (c) sets the first three parameters to 0.3, 0.3, 0.6, and the remaining 0.1; (d) is a comparison chart with increasing the delay weight of 0.7 and decreasing the packet loss weight of 0.1; (e) A comparison chart with the first three parameters set to 0.1, 0.3, 0.7; (f) a comparison chart with the first three parameters set to 0.7, 0.1, 0.3.}
	\label{fig11:weight factor settings}
\end{figure*}

To evaluate the efficacy of these parameter settings, we first set all parameter weights to 1 and conducted comparative experiments. As shown in Fig. \ref{subfig:fig11a}, weights of [0.7, 0.3, 0.1, 0.1, 0.1, 0.1, 0.1] achieve better convergence and yield higher reward values compared to setting all parameter weights to 1.

Then, to test the influence of the last four parameters on multicast tree construction, we set the parameters to [0.1, 0.1, 0.1, 0.7, 0.7, 0.6, 0.1] and compared the results with those of the initial weight ratios we set. As shown in Fig. \ref{subfig:fig11b}, the initial weight ratios we set show more stable convergence, so we set the weights of the last four parameters smaller.

Next, we tested changing the weight values of the first three parameters to [0.3, 0.3, 0.6, 0.1, 0.1, 0.1, 0.1]. As shown in Fig. \ref{subfig:fig11c}, although the convergence is relatively stable in both cases, the initial weight ratios we set yielded higher reward values and faster convergence.

Finally, to further verify the influence of the three main factors (remaining bandwidth, delay, and packet loss rate), we increased the weight of the delay parameter, setting the weights to [0.3, 0.7, 0.1, 0.1, 0.1, 0.1, 0.1]. As shown in Fig. \ref{subfig:fig11d}, the obtained reward value decreased slightly.

Similarly, we increased the weight of the packet loss rate and decreased the weight of the remaining bandwidth, setting the weights to [0.1, 0.3, 0.7, 0.1, 0.1, 0.1, 0.1]. As shown in Fig. \ref{subfig:fig11e}, tthe converged reward value achieved under this setting was closer to that achieved with the initial weight values, but the initial weight values we set are still better.

To further compare the importance of the delay and packet loss rate, we then swapped their weights, setting them to [0.7, 0.1, 0.3, 0.1, 0.1, 0.1, 0.1]. As shown in Fig. \ref{subfig:fig11f}, the initial weight values still yielded a higher reward value.

Through multiple adjustments of the parameter weights, the results consistently showed that the initial weights [0.7,0.3,0.1,0.1,0.1,0.1,0.1] offer the best convergence behavior and the highest reward values. These findings validate the reasonableness and accuracy of our initial weight setting.

\begin{figure}[htb]
	\centering
	\includegraphics[width=0.45\textwidth]{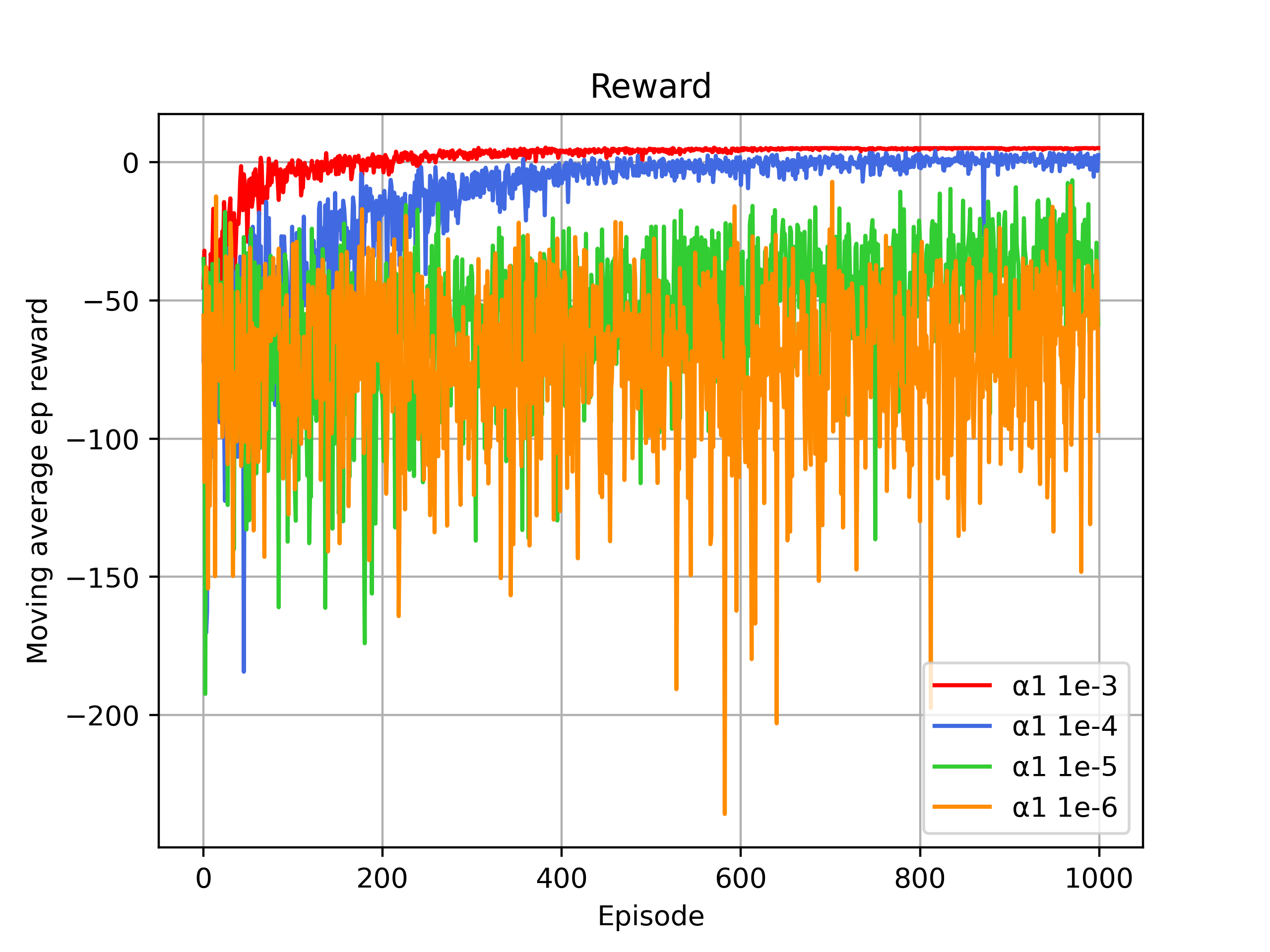}
	\caption{\raggedright Learning rate $ \alpha_1 $}
	\label{fig12}
\end{figure}

The learning rate is a hyperparameter that controls the speed at which a neural network adjusts its weights based on the loss gradient, directly impacting how quickly an agent can converge to the optimal value. Generally, a higher learning rate leads to faster learning of the neural network, while a lower learning rate may cause the model to become trapped in a local optimum. However, if the learning rate is too high, this can cause oscillation in the loss function during the parameter optimization process, leading to failure to converge. Therefore, setting a proper learning rate is crucial. The algorithm used in this paper is A2C, which involves two neural networks, the actor network and the critic network. To optimize the learning rates of the actor network $ (\alpha_1) $ and the critic network $ (\alpha_2) $, we fixed one learning rate and adjusted the other. First, we set $ \alpha_2=3e-3 $ and adjusted $ \alpha_1 $. The results are shown in Fig. \ref{fig12}.

\begin{figure}[htb]
	\centering
	\includegraphics[width=0.45\textwidth]{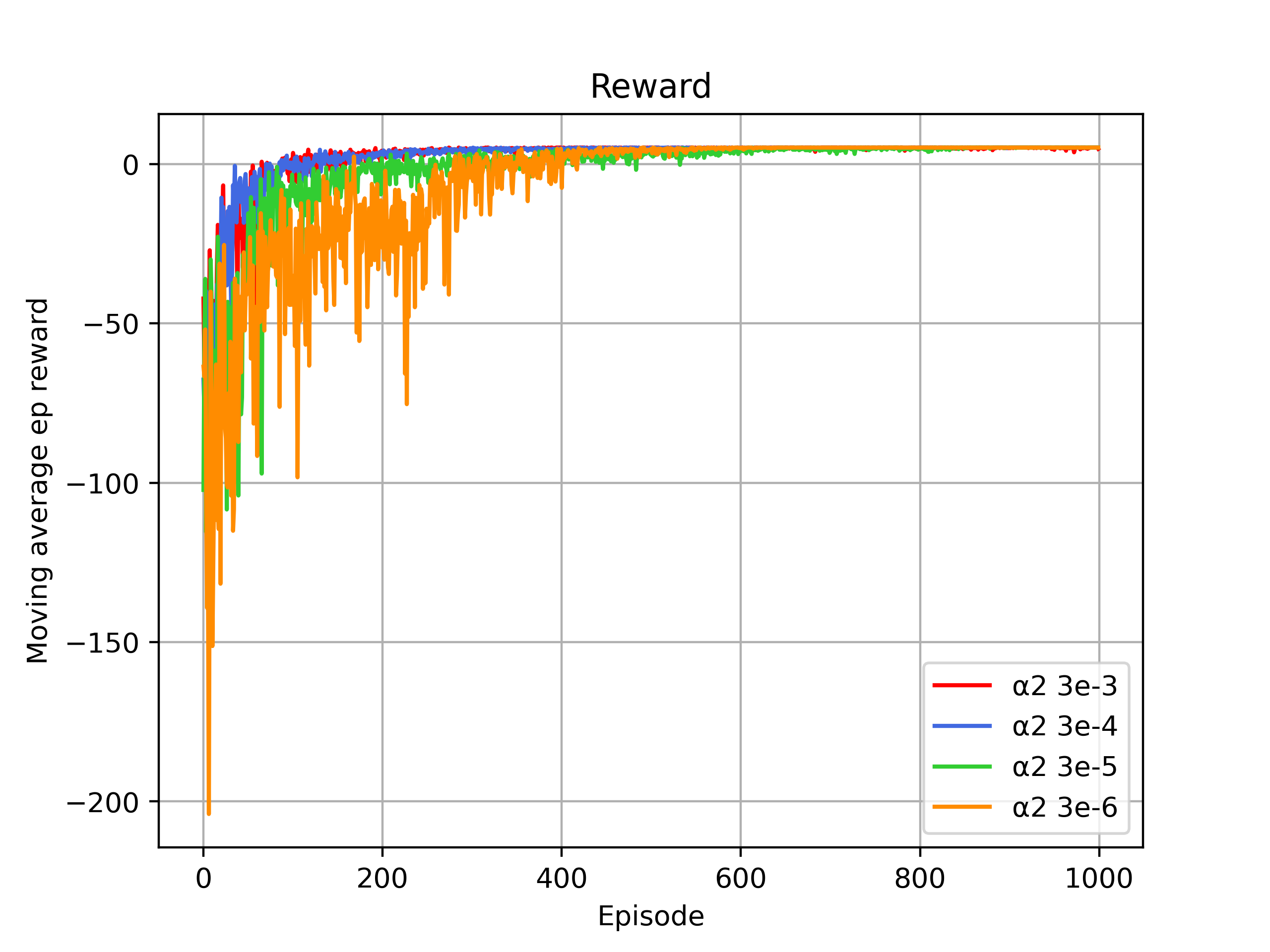}
	\caption{\raggedright Learning rate $ \alpha_2 $}
	\label{fig13}
\end{figure}

Based on the results of adjusting $ \alpha_1 $, it was found that when $ \alpha_1 $ was set to $ 1e-5 $ or $ 1e-6 $, the learning rate was too low for the network to converge easily. When $ \alpha_1 $ was set to 1e-3 or 1e-4, the reward value converged, but the convergence effect and reward value obtained with $ \alpha_1=1e-3 $ were the best. Then, $ \alpha_1 $ was fixed while $ \alpha_2 $ was adjusted, as shown in Fig. \ref{fig13}. The reward value converged under all tested values of $ \alpha_2 $, but when $ \alpha_2 $ was set to $ 3e-3 $ or $ 3e-4 $, he convergence speed was faster.

Based on the results of the above two comparative experiments, we set $ \alpha_1=1e-3 $ and $ \alpha_2=3e-3 $.

\begin{figure}[htb]
	\centering
	\includegraphics[width=0.45\textwidth]{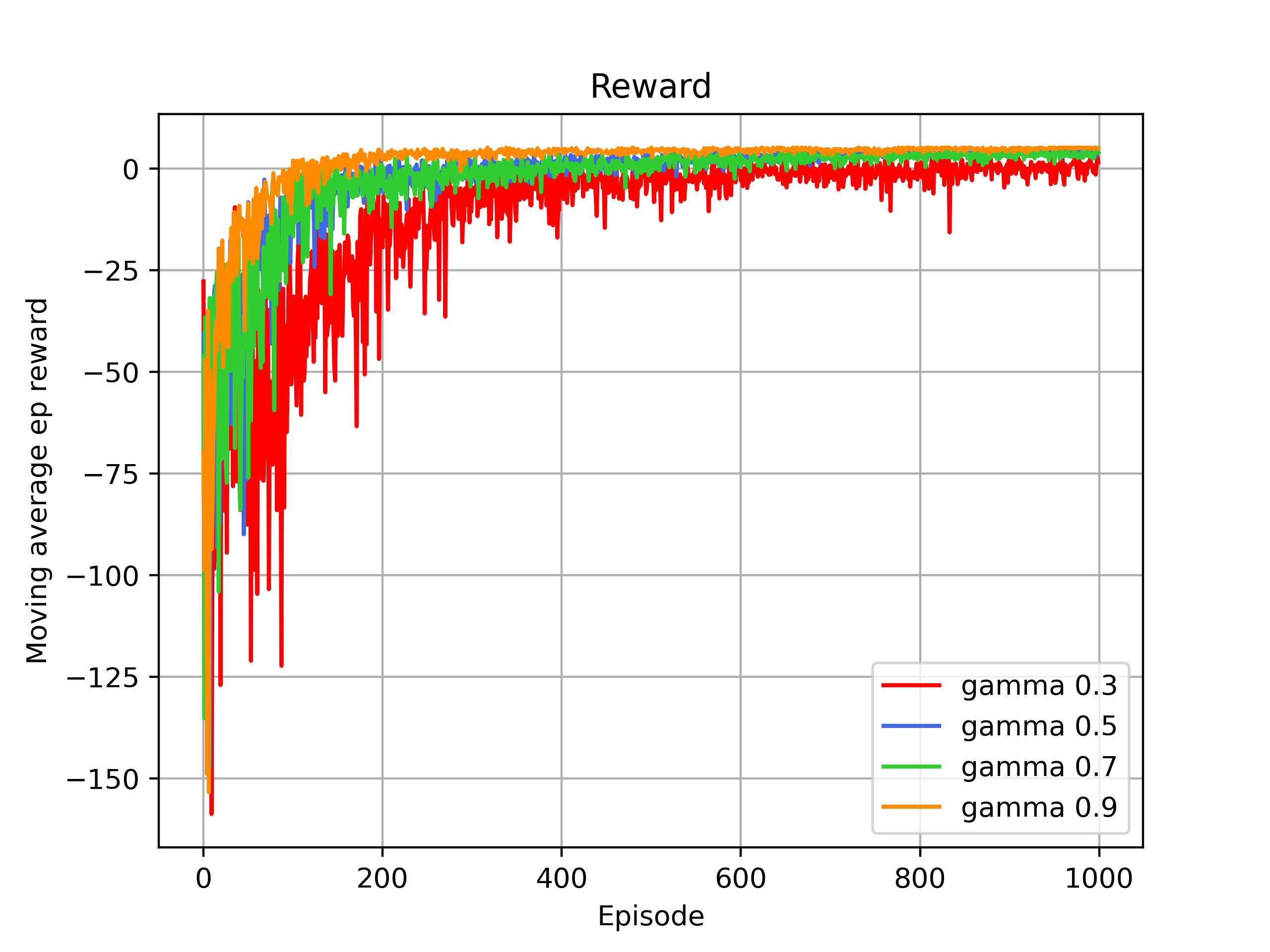}
	\caption{\raggedright Discount factor}
	\label{fig14}
\end{figure}

Based on the characteristics of the Markov process, we set a reward discount factor that discounts the rewards obtained in the future, with a greater discount for rewards from further ahead. This is because we wish to prioritize the current reward and avoid infinite rewards. By adjusting the discount factor (another hyperparameter), we can obtain intelligent agents with different performance, as shown in Fig. \ref{fig14}.

According to the results shown in Fig.\ref{fig14}, setting the discount factor to 0.9 was found to yield the best performance.

\begin{figure}[htb]
	\centering
	\includegraphics[width=0.45\textwidth]{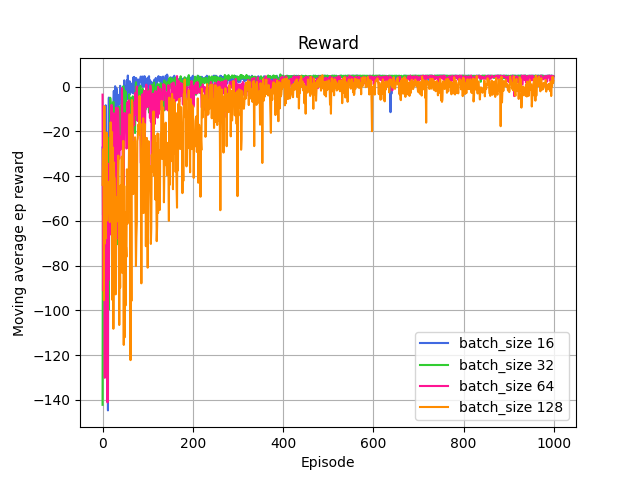}
	\caption{\raggedright batch size}
	\label{fig15}
\end{figure}

The purpose of the batch size hyperparameter is to control the number of samples selected by the model during each training iteration, which can affect the degree and speed of model optimization. From Fig. \ref{fig15}, we can see that the convergence situation is similar with batch sizes of 16, 32, and 64, whereas the performance obtained with a batch size of 128 is the worst.

\begin{figure}[htb]
	\centering
	\includegraphics[width=0.45\textwidth]{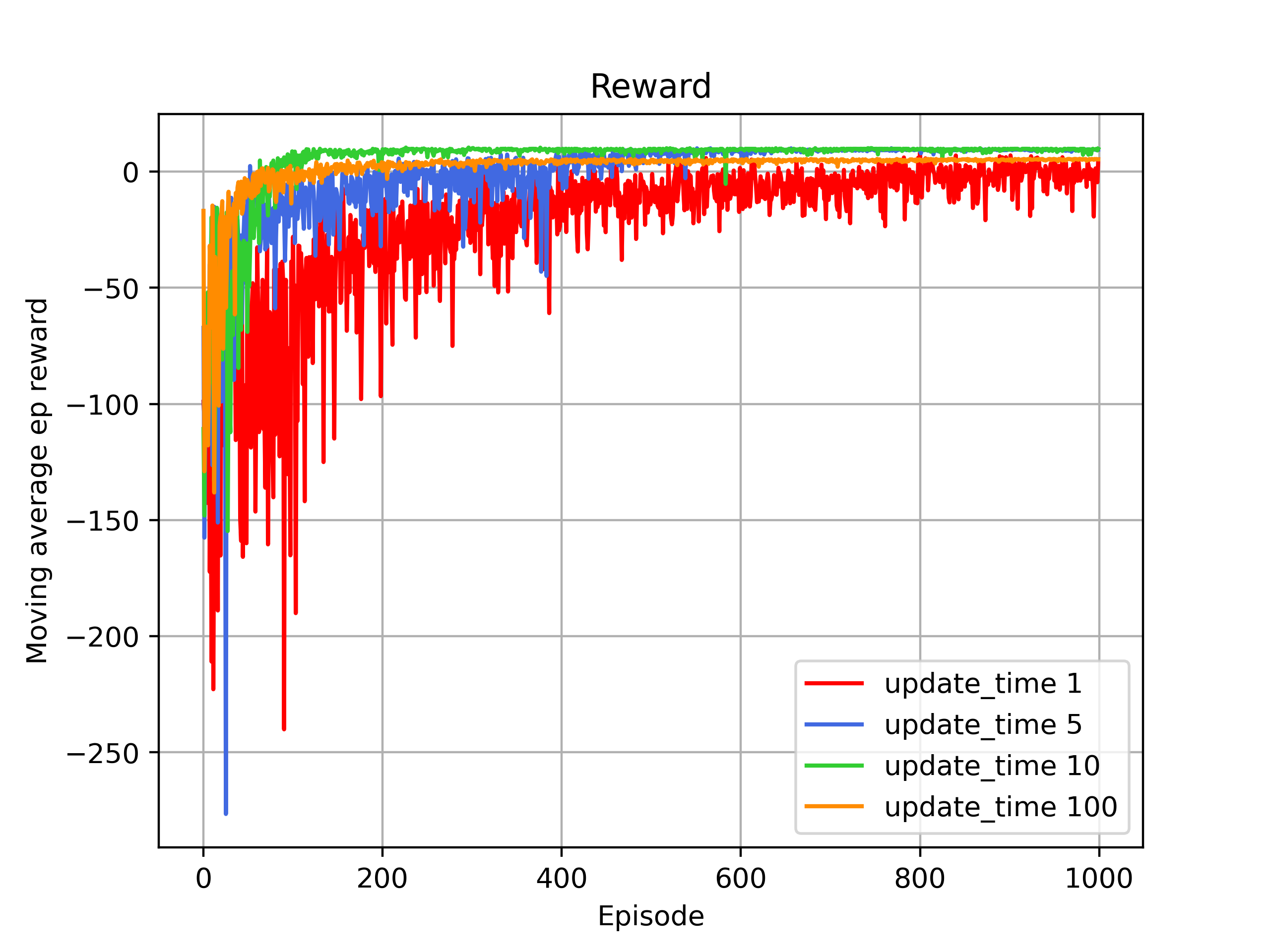}
	\caption{\raggedright update frequency}
	\label{fig16}
\end{figure}

Experiments on setting the update frequency of the neural networks of the intelligent agents, as shown in Fig. \ref{fig16}, were conducted by adjusting the $ update\_time $ parameter to 1, 10, 100, and 1000. When this parameter was set to 1000, the intelligent agents were unable to reach the terminal state, so an additional parameter adjustment experiment with $ update\_time $ set to 5 was added. The results in this figure show that the best convergence effect and the highest reward value were obtained when $ update\_time $ was set to 10.

\begin{figure}[htb]
	\centering
	\includegraphics[width=0.45\textwidth]{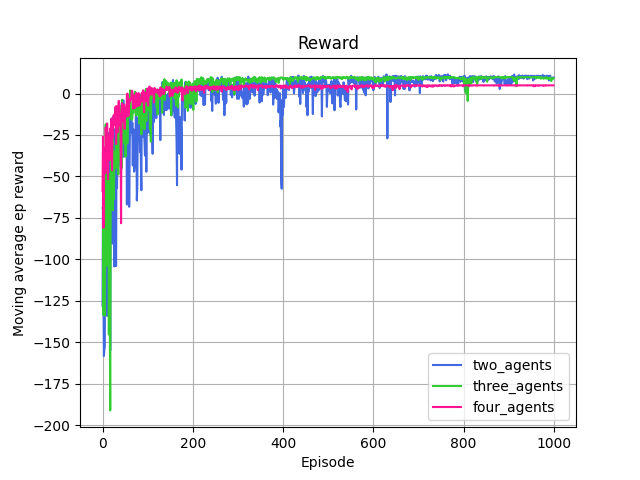}
	\caption{\raggedright Number of agents}
	\label{fig17}
\end{figure}

Multi-agent reinforcement learning is used in this paper to find the optimal policy; therefore, another very important hyperparameter is the number of agents. The convergence speed and the final converged reward value are different with different numbers of agents. It is not generally true that more agents are better. In fact, the larger the number of agents, the harder it is for the multi-agent system to converge. In Fig. \ref{fig17}, the experimental results show that when the number of agents is set to 3, the convergence of the reward value is the best.

\subsection{Comparative experiment}
\label{sec:5.4}
Reinforcement learning algorithms can be divided into two main categories: value-based and policy-based algorithms. Value-based algorithms perform well in discrete action spaces but struggle with continuous and high-dimensional action spaces. Additionally, decisions made using value-based methods are deterministic, although action selection probabilities can be set. Policy-based algorithms perform well in continuous action spaces and high-dimensional discrete action spaces but struggle with discrete action spaces. Each approach has its own advantages and disadvantages.

The MADRL-MR algorithm presented in this paper combines value-based and policy-based methods using an AC algorithm. The agent selects actions based on its policy, and the critic uses a value function to assign a value to each action taken. This approach facilitates faster learning on top of an existing policy. Although we have designed the action space to be discrete, we can use a probability distribution to select a specific action in a continuous action space. This approach can improve the randomness of the agents' actions and allow for a wider range of exploration.

\begin{figure}[htb]
	\centering
	\includegraphics[width=0.45\textwidth]{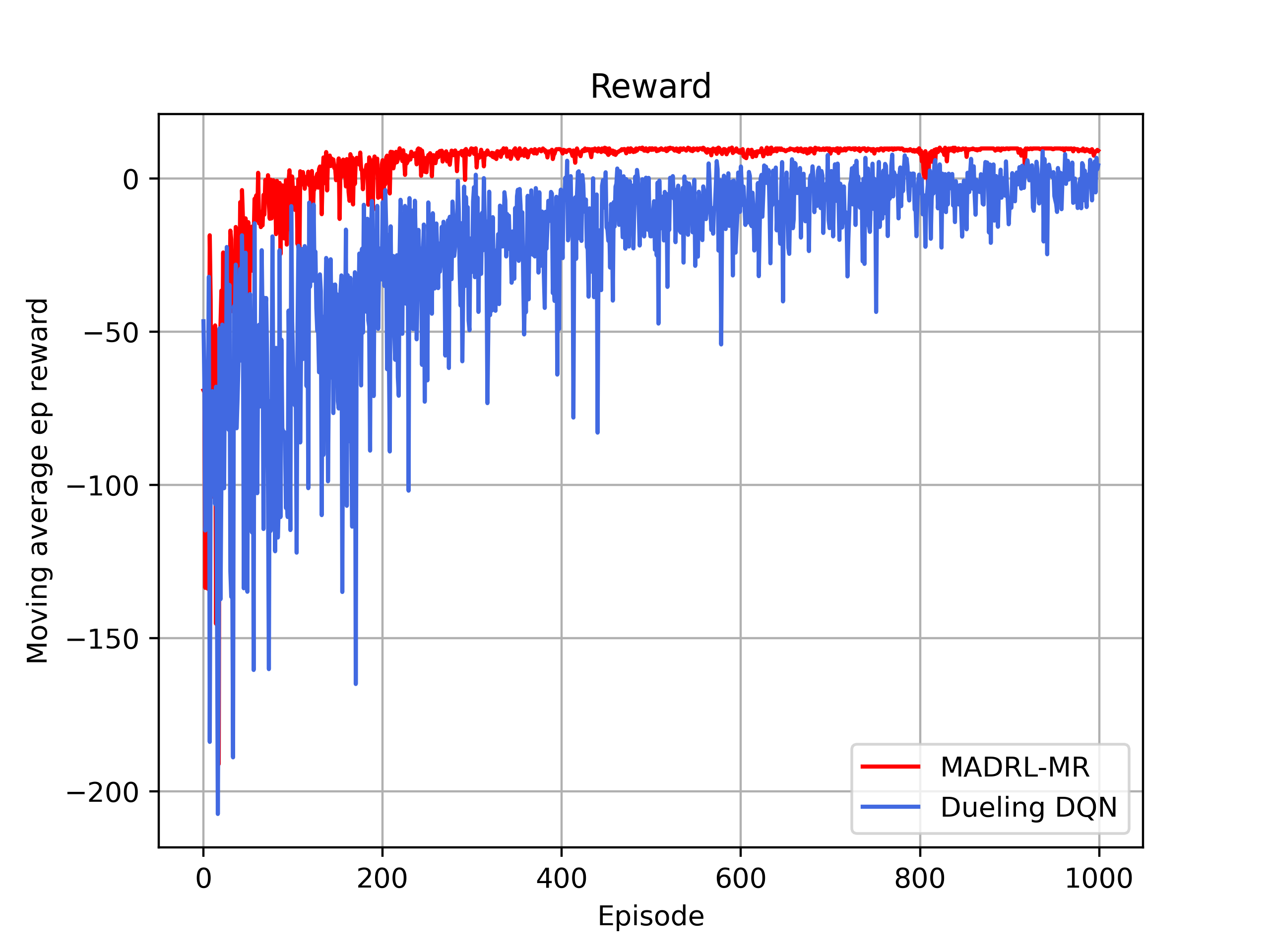}
	\caption{\raggedright Comparison between MADRL-MR and Dueling DQN}
	\label{fig18}
\end{figure}

To evaluate the performance of MADRL-MR algorithm, we compared it with deep reinforcement learning based on value function using double dueling deep Q network (DQN) and priority experience replay (PER). As shown in Fig. \ref{fig18}, the MADRL-MR algorithm achieved higher reward values, faster convergence, and better performance throughout the entire training process compared to double dueling DQN.

\begin{figure*}[htb]
	\centering
	\begin{minipage}{0.45\textwidth}
		\subfigure[throughput]{
			\centering
			\includegraphics[width=1\textwidth]{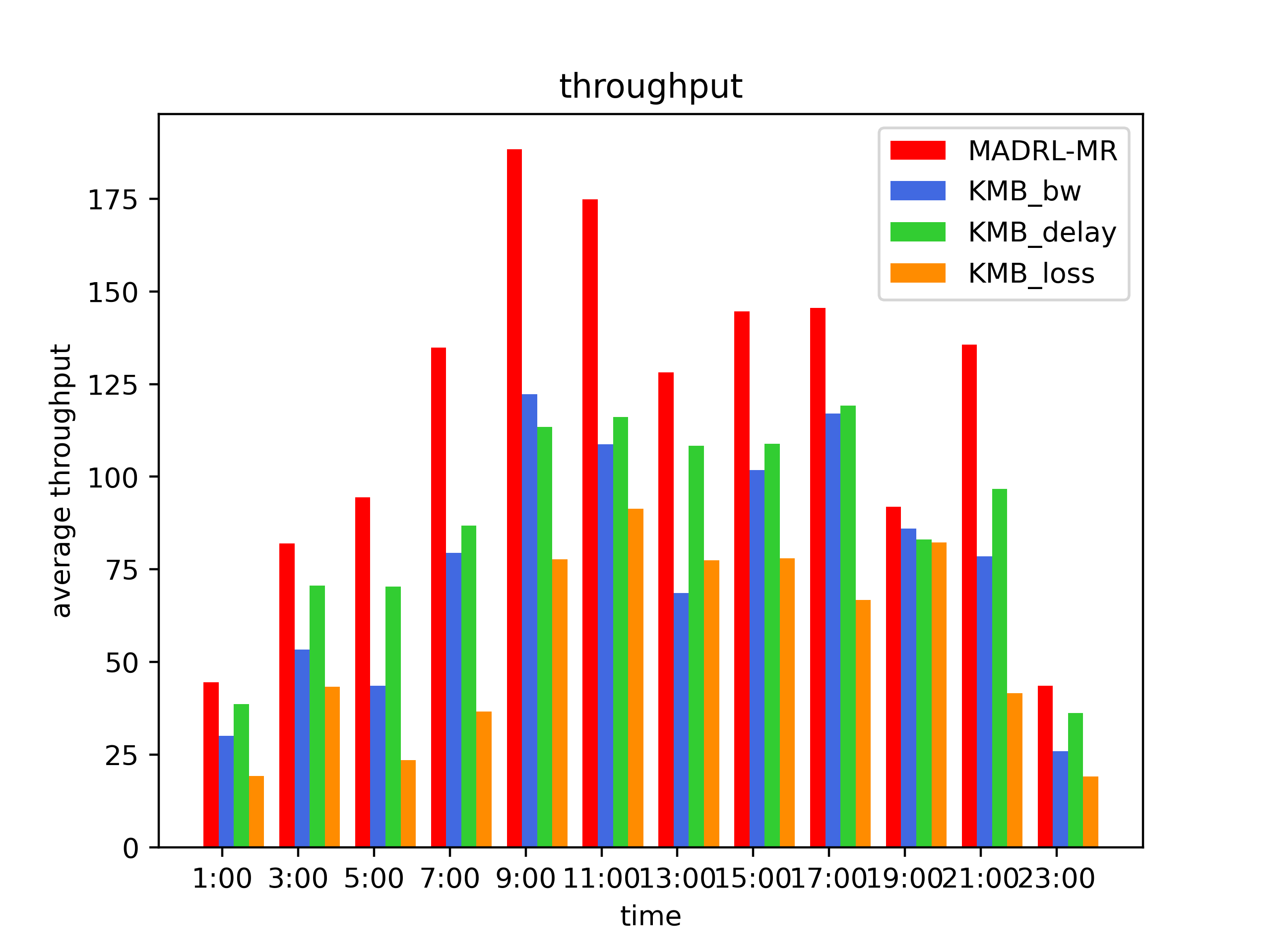}\hspace{-15pt}
			\label{subfig:fig19a}
		}
	\end{minipage}
	\begin{minipage}{0.45\textwidth}
		\subfigure[delay]{
			\centering
			\includegraphics[width=1\textwidth]{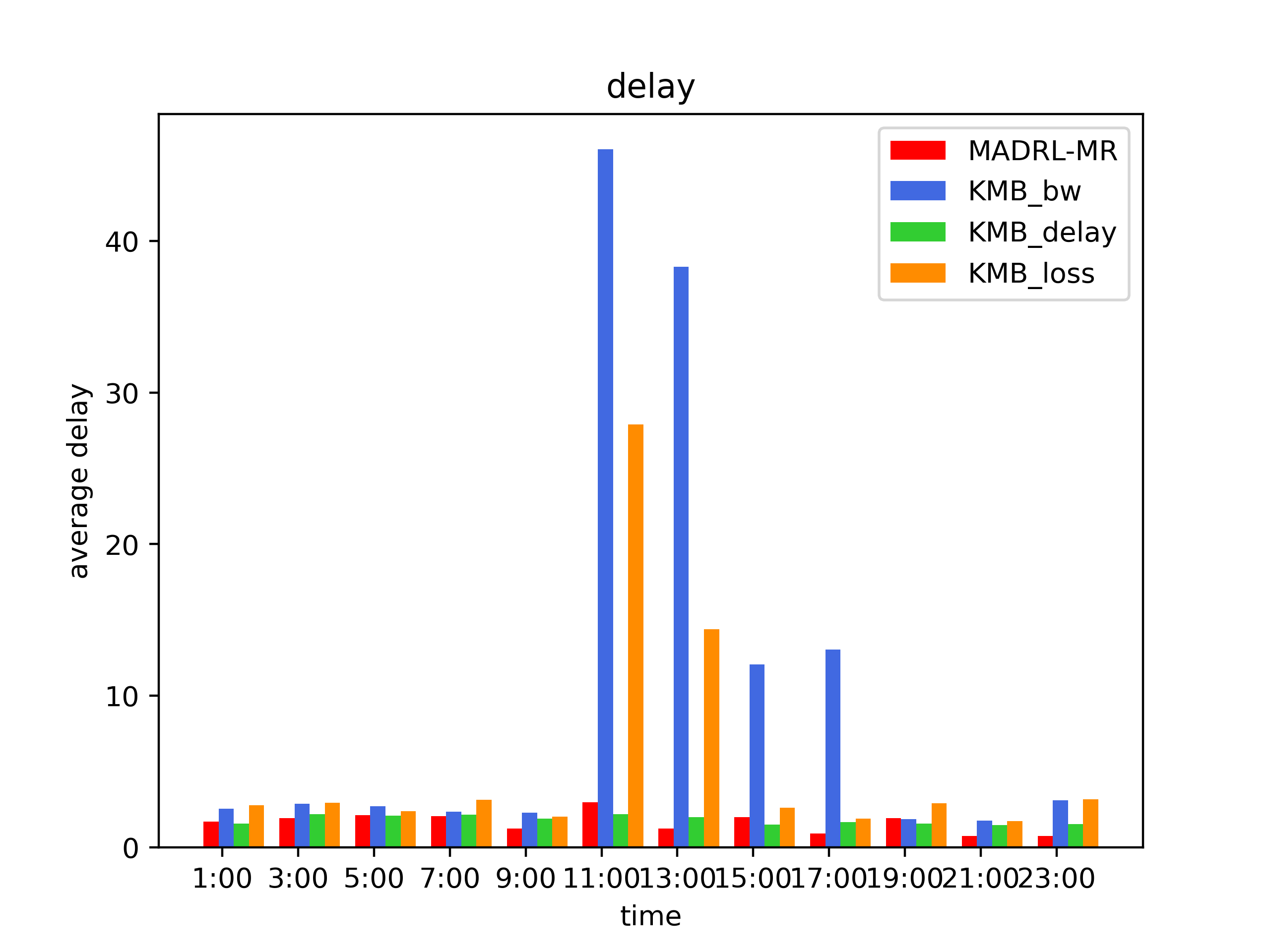}\hspace{-15pt}
			\label{subfig:fig19b}
		}
	\end{minipage}
	\begin{minipage}{0.45\textwidth}
		\subfigure[loss]{
			\centering
			\includegraphics[width=1\textwidth]{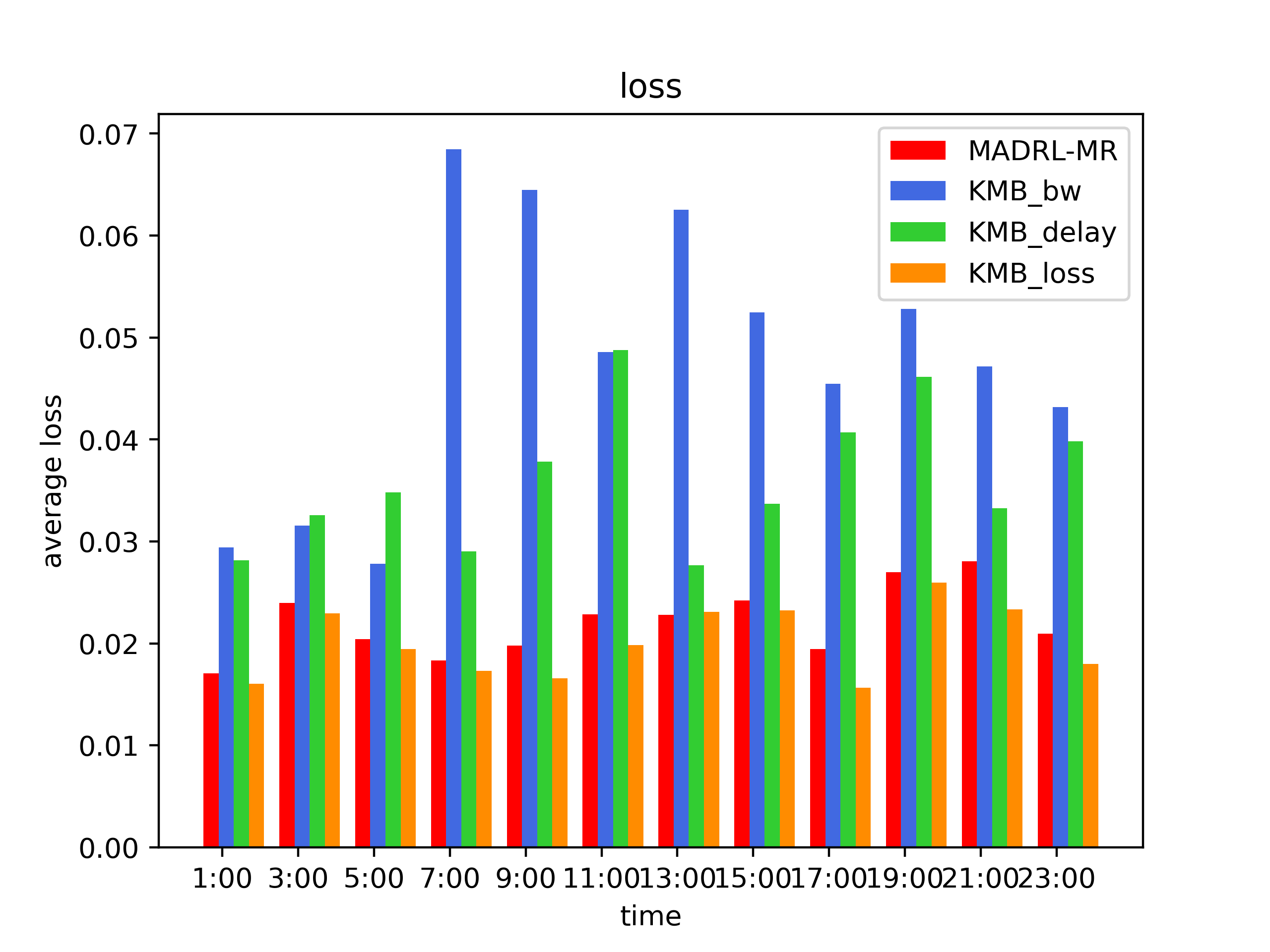}\hspace{-15pt}
			\label{subfig:fig19c}
		}
	\end{minipage}
	\begin{minipage}{0.45\textwidth}
		\subfigure[bandwidth]{
			\centering
			\includegraphics[width=1\textwidth]{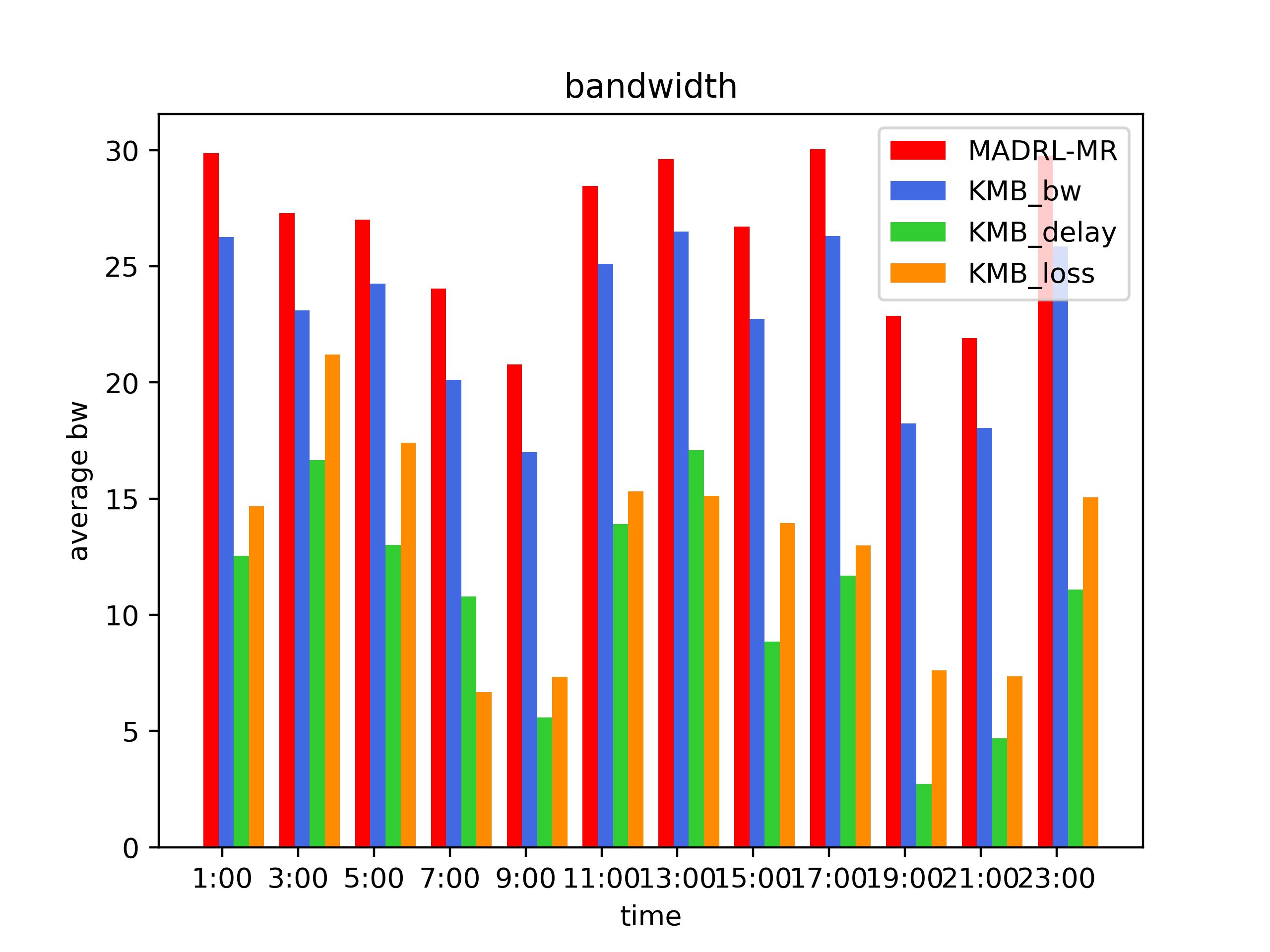}\hspace{-15pt}
			\label{subfig:fig19d}
		}
	\end{minipage}
	\begin{minipage}{0.45\textwidth}
		\subfigure[length]{
			\centering
			\includegraphics[width=1\textwidth]{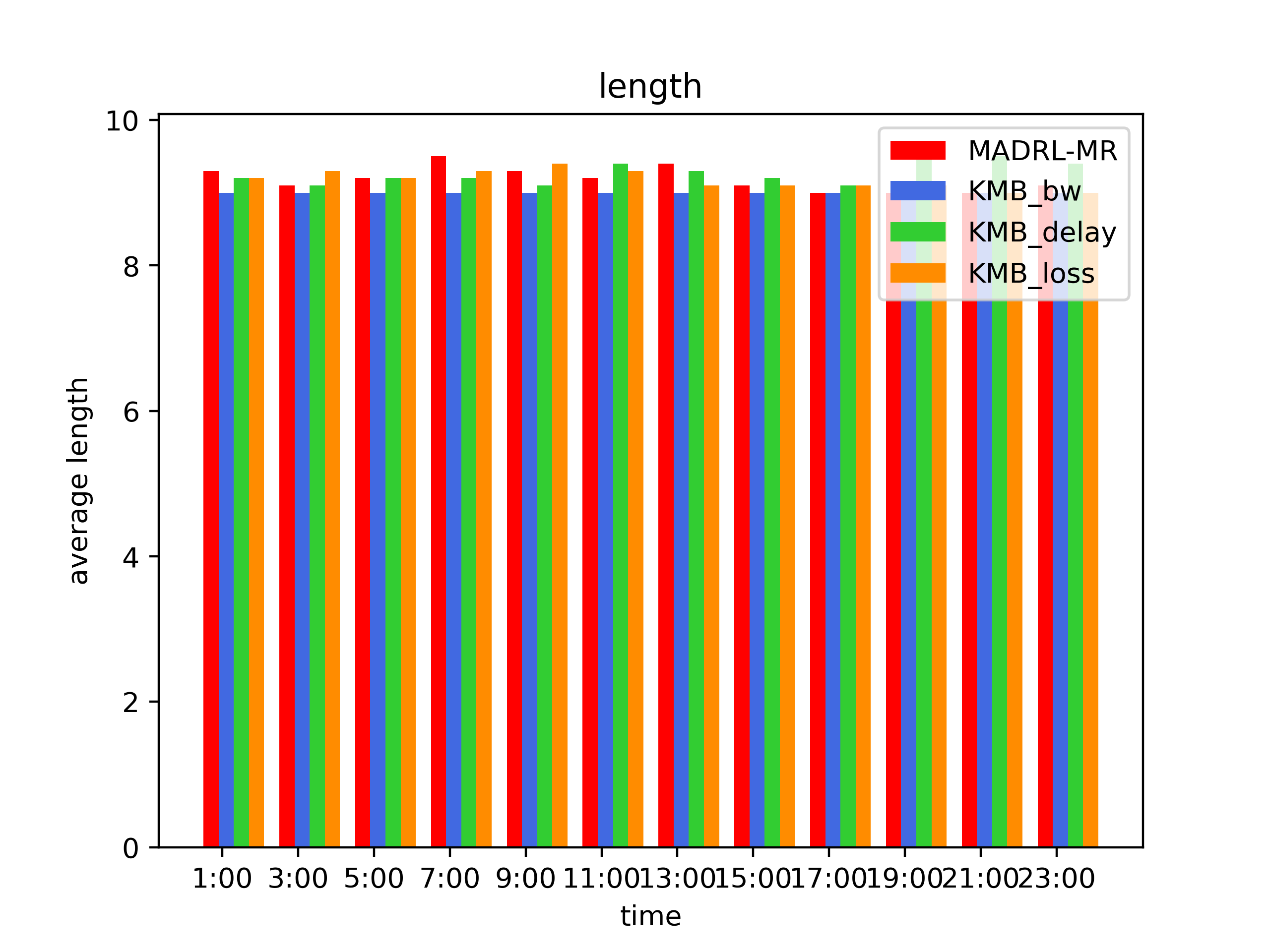}\hspace{-15pt}
			\label{subfig:fig19e}
		}
	\end{minipage}	
	\begin{minipage}{0.45\textwidth}
		\subfigure[distance]{
			\centering
			\includegraphics[width=1\textwidth]{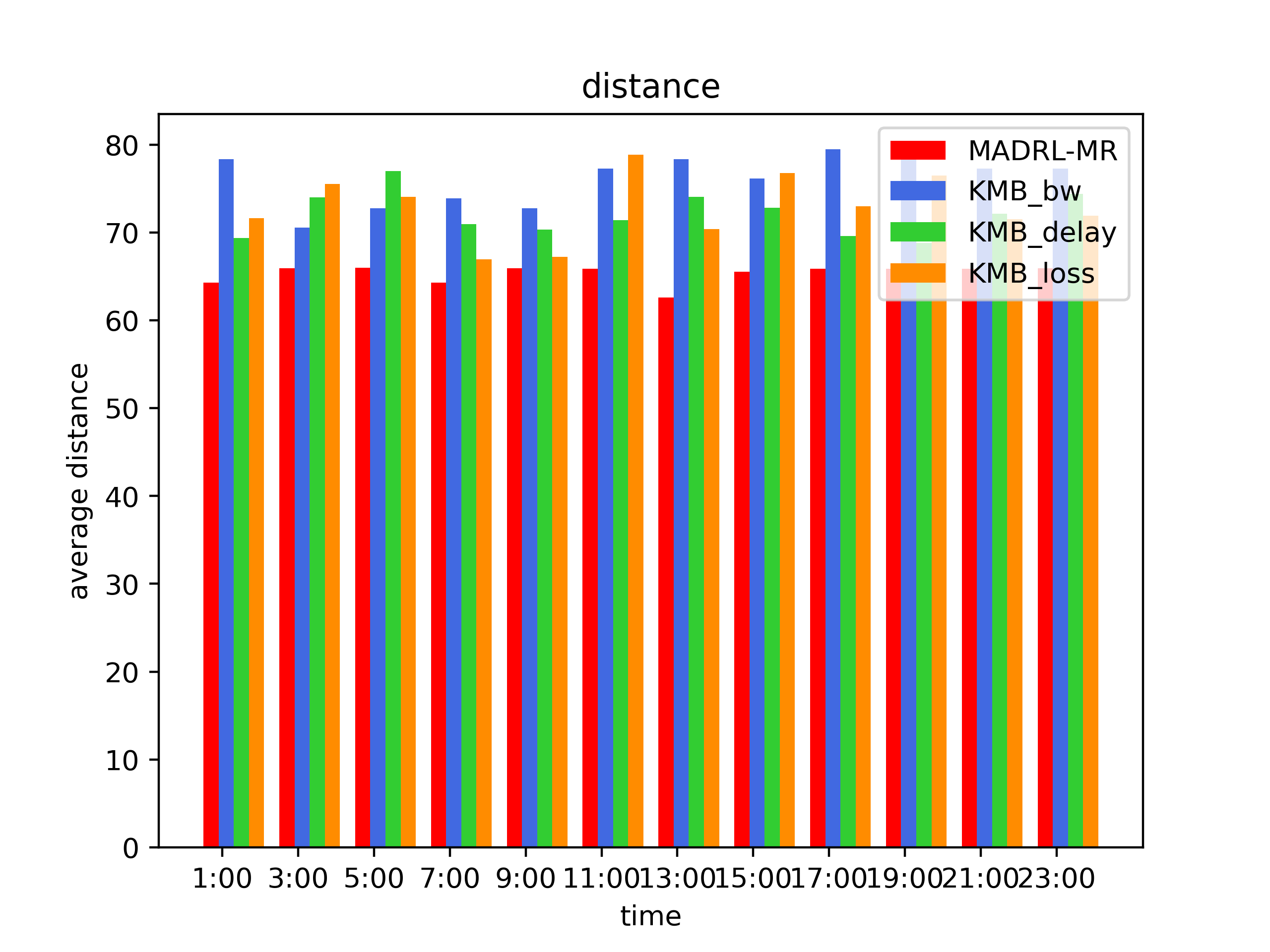}\hspace{-15pt}
			\label{subfig:fig19f}
		}
	\end{minipage}	
	\caption{The performance comparison of MADRL-MR with $ KMB_{bw} $, $ KMB_{delay} $, and $ KMB_{loss} $, (a) is to compare network throughput; (b) Compare the average link delay; (c) Compare the average link packet loss rate; (d) Compare the average link remaining bandwidth of multicast trees; (e) Compare the average length of multicast trees; (f) Compare the average distance of each link in a multicast tree.}
	\label{fig19}
\end{figure*}

We also compared the results of our algorithm with the experimental results of constructing Steiner trees using the classic KMB algorithm. To demonstrate the influence of the network link parameters on multicast tree construction, we implemented three versions of the KMB algorithm using the residual bandwidth, delay, and packet loss rate as weights. We used the average throughput, delay, packet loss rate, residual bandwidth, tree length, and average distance between wireless APs in the multicast tree as performance evaluation indicators. The results are shown in Fig. \ref{fig19}.

Fig. \ref{subfig:fig19a} presents the network throughput experiment comparing MADRL-MR with the KMB algorithm using the bandwidth, delay, and packet loss rate as weights. It can be observed that as time progresses and the simulated network traffic grows, the network throughput under the proposed intelligent multicast routing algorithm is significantly higher than those under $ KMB_{bw} $, $ KMB_{delay} $, and $ KMB_{loss} $. On average, it is 58.71\% higher than that under $ KMB_{bw} $ and 31.8\% higher than that under $ KMB_{delay} $.

Fig. \ref{subfig:fig19b} compares the average link delay of the multicast trees constructed by MADRL-MR, $ KMB_{bw} $, $ KMB_{delay} $, and $ KMB_{loss} $. As the network traffic increases, the average link delay of MADRL-MR is 53.52\% and 48.53\% lower than those of $ KMB_{bw} $ and $ KMB_{loss} $ respectively, and is close to the value for $ KMB_{delay} $. This indicates that MADRL-MR achieves good performance in terms of the average link delay.

Fig. \ref{subfig:fig19c} ccompares the average packet loss rates on the links of the multicast trees constructed by MADRL-MR, $ KMB_{bw} $, $ KMB_{delay} $, and $ KMB_{loss} $. MADRL-MR performs slightly worse than $ KMB_{loss} $ in terms of the average link packet loss rate, although the values of the two are very close. However, MADRL-MR outperforms $ KMB_{bw} $ and $ KMB_{delay} $ by 50.32\% and 37.3\% on average, respectively, indicating that MADRL-MR generally has a lower packet loss rate.

Fig. \ref{subfig:fig19d} compares the average link bandwidths of the multicast trees constructed by MADRL-MR, $ KMB_{bw} $, $ KMB_{delay} $, and $ KMB_{loss} $. MADRL-MR performs significantly better than $ KMB_{delay} $ and $ KMB_{loss} $ in terms of the average link bandwidth and exhibits an average improvement of 16.96\% compared to $ KMB_{bw} $.

Fig. \ref{subfig:fig19e} compares the average lengths of the multicast trees constructed by MADRL-MR, $ KMB_{bw} $, $ KMB_{delay} $, and $ KMB_{loss} $. The results show that the multicast tree constructed by MADRL-MR is longer on average than those constructed by the other three algorithms, reflecting the fact that our algorithm considers more parameters when constructing the multicast tree and considers more nodes when selecting nodes to join the multicast paths.

Fig. \ref{subfig:fig19f} compares the average distances between wireless AP nodes in the multicast trees constructed by MADRL-MR, $ KMB_{bw} $, $ KMB_{delay} $, and $ KMB_{loss} $. Despite the longer average multicast tree length of MADRL-MR shown in Fig. \ref{subfig:fig19e}, the distance between AP nodes does not show the same trend. As seen inthe distance between AP nodes does not show the same trend. As seen in Fig. \ref{subfig:fig19f}, the MADRL-MR algorithm constructs multicast trees with a shorter average distance between AP nodes than $ KMB_{bw} $, $ KMB_{delay} $, and $ KMB_{loss} $, indicating that the proposed algorithm considers the distance between wireless AP nodes and achieves good results.

\section{Conclusion}
\label{sec:Conclusion}
In this paper, we have introduced MADRL-MR, an intelligent multicast routing method based on multi-agent deep reinforcement learning in an SDWN environment. First, we addressed the issues of traditional wireless networks, such as the difficulty of controlling and maintaining nodes and the tight coupling of data forwarding and logic control in traditional network devices, making it difficult to achieve compatibility with other devices and software. To overcome these issues, we chose to utilize the decoupling of forwarding and control and the global perception capabilities in SDWN. Second, traditional multicast routing algorithms cannot effectively use the link information of the entire network to construct a multicast tree. Moreover, in deep reinforcement learning, multicast tree construction by a single agent has a slow convergence rate, leading to difficulty in responding quickly to dynamic network changes. MADRL-MR effectively utilizes network link information and rapidly constructs the optimal multicast tree through mutual cooperation among multiple agents. Finally, to speed up multiagent training, the use of transfer learning techniques was proposed to accelerate the convergence rate of the agents.

In MADRL-MR, the design of the agents is based on the traffic matrix and the process of multicast tree construction. The state space is designed based on these factors. The design of the action space is different from that in other algorithms that use the k-paths approach because in k-paths, the paths are fixed, and the optimal path is chosen from among these fixed paths; however, the fixed path chosen in this way is not guaranteed to be the best. Therefore, a novel next-hop method is adopted instead to design the action space in this article. The agents explore and gradually construct the optimal multicast routes. For this purpose, a reward function is designed and calculated based on the traffic matrix of the network links.

The results of a large number of comparative experiments show that the proposed MADRL-MR algorithm offers better performance than three versions of the classic KMB algorithm implemented using the residual bandwidth, delay, and packet loss rate as weights. Additionally, in a dynamic network in which the traffic is changing in real time, MADRL-MR can quickly deploy multicast routing solutions.

Furthermore, with the development and promotion of SDWN technology, the size of networks is becoming increasingly larger. A single controller will not be able to meet the needs of large-scale SDWN networks. Therefore, in the future, we will consider designing an intelligent multicast routing algorithm based on multi-agent deep reinforcement learning under multi-controller SDWN scenarios.

\section*{Acknowledgment}
The author would like to thank the anonymous reviewers for their helpful comments.

\EOD

\end{document}